\documentclass[twoside,leqno,twocolumn]{article}  
\usepackage{ltexpprt} 

\usepackage{times}
\usepackage{epsfig}
\usepackage{color}
\usepackage{url}
\usepackage{float}

\usepackage{amsmath}
\usepackage{amssymb}
\usepackage{balance}

\usepackage{algorithm}
\usepackage{algorithmic}

\usepackage{hyperref}
\hypersetup{colorlinks,urlcolor=blue,citecolor=blue,linkcolor=blue}

\newcommand{\hide}[1]{}

\newcommand{\bit}{\begin{itemize}}
\newcommand{\eit}{\end{itemize}}

\newcommand{\beq}{\begin{equation}}
\newcommand{\eeq}{\end{equation}}

\newcommand{\select}{{\small \textsf{SELECT}}} 
\newcommand{\selectv}{{\small \textsf{SelectV}}}
\newcommand{\selecth}{{\small \textsf{SelectH}}}
\newcommand{\fulle}{{\small \textsf{Full}}}
\newcommand{\dive}{{\small \textsf{DivE}}}
\newcommand{\rande}{{\small \textsf{RandE}}}
\newcommand{\algorithms}{{\small \textsf{Algorithms}}}
\newcommand{\ebedwin}{{\small {EBED (win)}}}
\newcommand{\ebedwout}{{\small {EBED (wout)}}}
\newcommand{\ebeduin}{{\small {EBED (uin)}}}
\newcommand{\ebeduout}{{\small {EBED (uout)}}}
\newcommand{\ebedwdeg}{{\small {EBED (wdeg)}}}
\newcommand{\ebedudeg}{{\small {EBED (udeg)}}}

\newcommand{\ptsadwin}{{\small {PTSAD (win)}}}
\newcommand{\ptsadwout}{{\small {PTSAD (wout)}}}
\newcommand{\ptsaduin}{{\small {PTSAD (uin)}}}
\newcommand{\ptsaduout}{{\small {PTSAD (uout)}}}
\newcommand{\ptsadwdeg}{{\small {PTSAD (wdeg)}}}
\newcommand{\ptsadudeg}{{\small {PTSAD (udeg)}}}

\newcommand{\spiritwin}{{\small {SPIRIT (win)}}}
\newcommand{\spiritwout}{{\small {SPIRIT (wout)}}}
\newcommand{\spirituin}{{\small {SPIRIT (uin)}}}
\newcommand{\spirituout}{{\small {SPIRIT (uout)}}}
\newcommand{\spiritwdeg}{{\small {SPIRIT (wdeg)}}}
\newcommand{\spiritudeg}{{\small {SPIRIT (udeg)}}}

\newcommand{\asedwin}{{\small {ASED (win)}}}
\newcommand{\asedwout}{{\small {ASED (wout)}}}
\newcommand{\aseduin}{{\small {ASED (uin)}}}
\newcommand{\aseduout}{{\small {ASED (uout)}}}
\newcommand{\asedwdeg}{{\small {ASED (wdeg)}}}
\newcommand{\asedudeg}{{\small {ASED (udeg)}}}

\newcommand{\maedwin}{{\small {MAED (win)}}}
\newcommand{\maedwout}{{\small {MAED (wout)}}}
\newcommand{\maeduin}{{\small {MAED (uin)}}}
\newcommand{\maeduout}{{\small {MAED (uout)}}}
\newcommand{\maedwdeg}{{\small {MAED (wdeg)}}}
\newcommand{\maedudeg}{{\small {MAED (udeg)}}}

\newcommand{\IR}{{\small {Inverse Rank}}}
\newcommand{\KY}{{\small {Kemeny-Young}}}
\newcommand{\rra}{{\small {RRA}}}
\newcommand{\uniavg}{{\small {Uni (avg)}}}
\newcommand{\unimax}{{\small {Uni (max)}}}
\newcommand{\mmavg}{{\small {MM (avg)}}}
\newcommand{\mmmax}{{\small {MM (max)}}}

\usepackage{bm}
\usepackage{adjustbox,lipsum}
\usepackage{multirow}

\usepackage{multibib}
\newcites{latex}{Appendix References}%

\title{{\huge Less is More: Building Selective Anomaly Ensembles} \\
with Application to Event Detection in Temporal Graphs}

\begin{document}


\author{
\makebox[130pt]{Shebuti Rayana}\\
\makebox[130pt]{Stony Brook University}\\
\makebox[130pt]{srayana@cs.stonybrook.edu}
\and
\makebox[130pt]{Leman Akoglu}\\
\makebox[130pt]{Stony Brook University}\\
\makebox[130pt]{leman@cs.stonybrook.edu}
}
\date{}
\maketitle

\begin{abstract}

Ensemble techniques for classification and clustering have long proven effective, yet anomaly ensembles have been barely studied.
In this work, we tap into this gap and propose a new ensemble approach for anomaly mining, with application to event detection in temporal graphs.
Our method aims to combine results from heterogeneous detectors with varying outputs, and leverage the evidence from multiple sources to yield better  performance.
However, trusting \textit{all} the results may deteriorate the overall ensemble accuracy,  as some detectors may fall short  and provide inaccurate results depending on the nature of the data in hand. 
This suggests that being \emph{selective} in which results to combine is vital in building effective ensembles---hence ``less is more''.

In this paper we propose \select; an ensemble approach for anomaly mining that 
employs novel techniques to automatically and systematically select the results to assemble in a fully unsupervised fashion. 
We apply our method to event detection in temporal graphs, where \select~successfully utilizes five base detectors and seven consensus methods under a unified ensemble framework.
We provide extensive quantitative evaluation of our approach on five real-world datasets (four with ground truth), including Enron email communications, New York Times news corpus, and World Cup 2014 Twitter news feed. Thanks to its selection mechanism, \select~yields superior performance compared to individual detectors alone, the full ensemble (naively combining all results), and an existing diversity-based ensemble. 

\end{abstract}

\section{Introduction}
\label{sec:intro}

Ensemble methods utilize multiple algorithms to obtain  better performance than the constituent algorithms alone and produce more robust results \cite{conf/mcs/Dietterich00}. 
Thanks to these advantages, a large body of research has been devoted to ensemble learning in classification 
\cite{journals/pami/HansenS90,preisach2007,journals/air/Rokach10,conf/wirn/ValentiniM02}  
and clustering \cite{conf/sdm/FernL08,books/crc/aggarwal13/GhoshA13,journals/inffus/HadjitodorovKT06,journals/pami/TopchyJP05}. 
On the other hand, building effective ensembles for anomaly detection has proven to be a challenging task \cite{journals/sigkdd/Aggarwal12,Zimek13Ensemble}.
A key challenge is
the lack of ground-truth; which makes it hard to measure detector accuracy and to accordingly select accurate detectors to combine, unlike in classification. Moreover, there exist no objective or `fitness' functions for anomaly mining, unlike in clustering.

Existing attempts for anomaly ensembles 
either combine outcomes from all the constituent detectors
\cite{conf/pakdd/GaoHZW12,conf/icdm/GaoT06,conf/sdm/KriegelKSZ11,conf/kdd/LazarevicK05},
or induce diversity among their detectors to increase the chance that they make independent errors \cite{conf/sdm/SchubertWZK12,conf/kdd/ZimekGCS13}.
However, as our prior work \cite{RayanaAkoglu14} suggests, neither of these strategies would work well in the presence of inaccurate detectors.
In particular, combining all, including inaccurate results would deteriorate the overall ensemble performance. Similarly, diversity-based ensembles would combine inaccurate results for the sake of diversity.

In this work, we tap into the gap between anomaly mining and ensemble methods, and propose \select, 
one of the first \textit{selective} ensemble approaches for anomaly detection.
As the name implies, the {key} property of our ensemble is its selection mechanism which carefully decides which results to combine from multiple different methods in the ensemble.
We summarize our contributions as follows.

\vspace{-0.1in}
\bit
\setlength{\itemsep}{-1.0\itemsep}
\item We identify and study the problem of building selective
anomaly ensembles in a fully unsupervised fashion.

\item We propose \select, a new ensemble approach for anomaly detection, which utilizes not only multiple heterogeneous detectors, but also various consensus methods under a unified ensemble framework.

\item \select~employs two novel unsupervised selection strategies that we design to choose the detector/consensus results to combine, which render the ensemble not only more robust but improve its performance further over its non-selective counterpart.

\item Our ensemble approach is general and flexible. It does not rely on specific data types, 
 and allows other detectors and consensus methods to be incorporated. 
\eit
\vspace{-0.1in}

We apply our ensemble approach to the event detection problem in temporal graphs, where \select~utilizes five heterogeneous event detection algorithms and seven different consensus methods.
Extensive evaluation on datasets with ground truth shows that \select~outperforms the average individual detector, the full ensemble that naively combines all results, as well as the diversity-based ensemble in \cite{conf/sdm/SchubertWZK12}.

\clearpage

\section{Background and Preliminaries}
\label{sec:problem}


\subsection{Event Detection Problem}

Temporal graphs change dynamically over time in which new nodes and edges arrive or existing nodes and edges disappear.
Many dynamic systems can be modeled as temporal graphs, such as computer, trading, transaction, and communication networks.

Event detection in temporal graph data
is the task of finding the points in time at which the graph structure notably differs from its past. 
These change points may correspond to significant events; such as critical state changes, anomalies, faults, intrusion, etc. depending on the application domain. Formally, the problem can be stated as follows.

%

\vspace{0.025in}
\noindent
{\bf Given} a sequence of graphs $\{G_1, G_2, \ldots, G_t, \ldots, G_T\}$;

\noindent
{\bf Find} time points $t'$ s.t. $G_{t'}$ differs significantly from $G_{t'-1}$.


\vspace{-0.075in}
\subsection{Motivation for Ensembles}

Several different methods have been proposed for the above problem, a survey of which is given in \cite{DBLP:journals/corr/AkogluTK14}. 
To date, however, there exists no single method that has been shown to outperform all the others. 
The lack of a winner technique is not a freak occurrence. In fact, it is unlikely that a given method could perform consistently well on different data of varying nature. 
Further, different techniques may identify different classes or types of anomalies depending on their particular formulation. 
This suggests that effectively \textit{combining} the results from various different detection methods (detectors from here onwards) could help improve the detection performance. 

\vspace{-0.075in}
\subsection{Motivation for Selective Ensembles}
\label{sec:selectmotiv}

Ensembles are expected to perform superior to their average constituent detector, however a naive ensemble  that trusts results from \textit{all} detectors may not work well. The reason is, some methods may not be as effective as desired depending on the nature of the data in hand, and fail to identify the anomalies of interest. As a result, combining accurate results with inaccurate ones may deteriorate the overall ensemble performance \cite{RayanaAkoglu14}.
This suggests that \emph{selecting} which detectors to assemble is a critical aspect of building robust ensembles---which implies that ``less is more''.

\begin{figure}[!t]
  \centerline{\psfig{figure=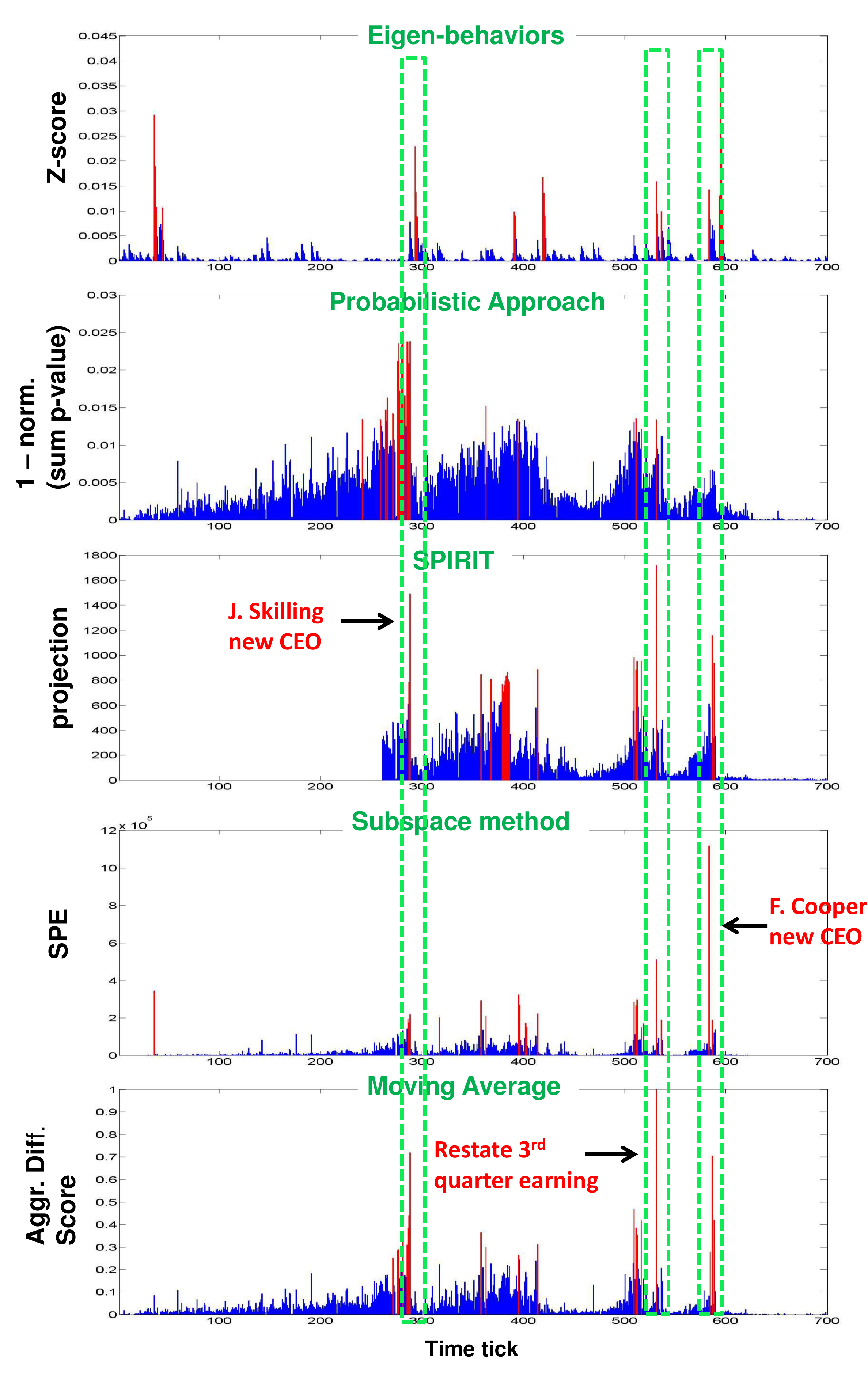,width=0.5\textwidth, height=4.25in} }
  \vspace{-0.2in}
	\caption{\small Anomaly scores from five detectors (rows) for the Enron Inc. time line. Red bars depict top 20 anomalous time points.}
	 \vspace{-0.25in}
	\label{fig:motiv}
\end{figure}

To illustrate the motivation for (selective) ensemble building further, consider the example in Figure \ref{fig:motiv}. The rows show the anomaly scores assigned by five different detectors to time points in the Enron Inc.'s time line. Notice that the scores are of varying nature and scale, due to different formulations of the detectors. We realize that the detectors mostly agree on the events that they detect; e.g., `J. Skilling new CEO'. On the other hand, they assign different magnitude of anomalousness to the time points; e.g., the top anomaly of methods varies. These suggest that combining the outcomes could help build improved ranking of the anomalies. Next notice the result provided by ``Probabilistic Approach'' which, while identifying one major event also detected by other detectors, fails to provide a reliable ranking for the rest; e.g., it scores many other time points higher than `F. Cooper new CEO'. As such, including this detector in the ensemble is likely to deteriorate the overall performance.

In summary, inspired by the success of classification and clustering ensembles and driven by the limited work on anomaly ensembles, 
we aim to systematically combine the strengths of accurate detectors while alleviating the weaknesses of the less accurate ones to build selective detection ensembles for anomaly mining.
While we build ensembles for the event detection problem in this paper, our approach is general and can directly be employed on a collection of detection methods for other anomaly mining problems.

\vspace{-0.075in}
\section{\select: \textsf{\small S}elective \textsf{\small E}nsemble \textsf{\small L}earning for 
anomaly det\textsf{\small ECT}ion --- Application to Event Detection}
\label{sec:method}


\subsection{Overview}

Our \select~approach takes the input data, in this case a sequence of graphs $\{G_1, \ldots, G_t, \ldots, G_T\}$, and outputs a rank list $R$ of objects, in this case of time points $1\leq t\leq T$, ranked from most to least anomalous. 

The main steps of \select~are given in Algorithm \ref{selectAlgo}. 
Step 1 employs (five) different event detection algorithms as base detectors of the ensemble.
Each detector has a specific and different measure to score the individual time points 
by anomalousness. As such, the ensemble embodies heterogeneous detectors.
As motivated earlier, Step 2 selects a subset of the detector results to assemble through a proposed selection strategy.
Step 3 then combines the selected results into a consensus. Besides several different event detection algorithms, there also exist various different consensus finding approaches. In spirit of building ensembles, \select~also leverages (seven) different consensus techniques to create intermediate aggregate results.
Similar to Step 2, 
Step 4 then selects a subset of the consensus results to assemble. 
Finally, Step 5 combines this subset into the final rank list of time points using inverse rank aggregation (Section \ref{ssec:consensus}). 

\vspace{-0.1in}
\begin{algorithm}[h]
\caption{\select}
\label{selectAlgo}
\begin{algorithmic}[1]
\REQUIRE Data: graph sequence $\{G_1, \ldots, G_t, \ldots, G_T\}$
\ENSURE Rank list of objects (time points) by anomaly 
\STATE Obtain results from (5) base detectors
\STATE Select set $E$ of detectors to assemble
\STATE Combine $E$ by (7) consensus techniques
\STATE Select set $C$ of consensus results to assemble
\STATE Combine $C$ into final rank list
\end{algorithmic}
\end{algorithm}
\vspace{-0.1in}


Different from prior works, ($i$) \select~is a \textit{two-phase} ensemble that not only leverages multiple detectors but also multiple consensus techniques, and ($ii$) it employs novel strategies to carefully select the ensemble components to assemble without any supervision, which outperform naive (no selection) and diversity-based selection (Section \ref{sec:result}).  Moreover, ($iii$) \select~is the first ensemble method for event detection in temporal graphs, although the same general framework as presented in Algorithm \ref{selectAlgo} can be deployed for other anomaly mining tasks, where the base detectors are replaced with a set of algorithms for the particular task at hand.

Next we fill in the details on the three main components of the proposed \select~ensemble. 
In particular, we describe the base detectors (Section \ref{ssec:base}), consensus techniques 
(Section \ref{ssec:consensus}), and the selection strategies (Section \ref{ssec:ensemble}).

%
%
%

\vspace{-0.1in}
\subsection{Base Detectors}
\label{ssec:base}

There exist various methods for the event detection problem in temporal graphs \cite{DBLP:journals/corr/AkogluTK14}.
In this work \select~employs five base detectors (Algorithm \ref{selectAlgo}, Line 1), while one can easily expand the ensemble with others: 
(1) eigen-behavior based event detection (EBED) from our prior work \cite{Akoglu2010army},
(2) probabilistic time series anomaly detection (PTSAD) we developed recently \cite{RayanaAkoglu14},
(3) Streaming Pattern DIscoveRy in multIple Time-Series (SPIRIT) by Papadimitriou \textit{et al.} \cite{Papadimitriou05streamingpattern},
(4) anomalous subspace based event detection (ASED) by Lakhina \textit{et al.} \cite{Lakhina04subspace}, and
(5) moving-average based event detection (MAED).
All methods extract graph-centric features (e.g., degree) for all nodes over time and detect events in multi-variate time series.
We provide brief descriptions of the methods in Appendix \ref{app:base} due to space limit.



\vspace{-0.1in}
\subsection{Consensus Finding}
\label{ssec:consensus}

Our ensemble consists of heterogeneous detectors.
That is, the detectors employ different anomaly scoring functions and hence their scores may vary in range and interpretation (see Figure \ref{fig:motiv}). Unifying these various outputs to find a consensus among detectors is an essential step toward building an ensemble.

A number of different consensus finding approaches have been proposed in the literature, which can be categorized into two, as rank based and score based aggregation methods.
Without choosing one over the other, we utilize seven well-established methods as  we describe below. 


\vspace{0.025in}
\noindent
\textbf{Rank based consensus.}
Rank based methods use the anomaly scores to order the data points (here, time points) into a rank list. This ranking makes the algorithm outputs comparable and facilitates combining them.
Merging multiple rank lists into a single ranking is known as rank aggregation, which has a rich history in theory of social choice 
and information retrieval \cite{Dwork2001}.
\select~employs three rank based consensus methods.
{\em Kemeny-Young}~\cite{Kemeny59} is a voting technique that uses preferential ballot and pair-wise comparison counts to combine multiple rank lists, in which the detectors are treated as voters and the points as the candidates they vote for.
\textit{Robust Rank Aggregation} (RRA)~\cite{Kolde2012} utilizes order statistics to compute the probability that a given ordering of ranks for a point across detectors is 
generated by the null model where the ranks are sampled from a uniform distribution.  The final ranking is done based on this probability, where more anomalous points receive a lower probability.
The third approach is based on \textit{Inverse Rank} aggregation, in which we score each point by $\frac{1}{r_i}$ where $r_i$ denotes its rank by detector $i$ and average these scores across detectors based on which we sort the points into a final rank list.


\vspace{0.025in}
\noindent
\textbf{Score based consensus.}
Rank-based aggregation provides a crude ordering of the data points, as it ignores the actual anomaly scores and their spacing. For instance, quite different rankings can yield equal performance in binary decision. 
Score-based aggregation approaches tackle the calibration of different anomaly scores and unify them
within a shared range.
\select~employs two score based consensus methods.
{\em Mixture Modeling} \cite{conf/icdm/GaoT06} converts the anomaly scores into probabilities by modeling them as sampled from a mixture of exponential (for inliers) and Gaussian (for outliers) distributions. 
{\em Unification} \cite{conf/sdm/KriegelKSZ11} also converts the scores into probability estimates through regularization, normalization, and scaling steps.
The probabilities are then comparable across detectors, which we aggregate by  
both $max$ and $avg$. This yields four score based methods.

\vspace{-0.05in}
\subsection{Ensemble Learning} 
\label{ssec:ensemble}

Given different base detectors and various consensus methods, the final task remains to utilize them under a unified ensemble framework. In this section, 
we discuss four different approaches for building anomaly ensembles.
These approaches differ in whether and how they select their ensemble components.

\vspace{-0.1in}
\subsubsection{Full ensemble}
The full ensemble selects all the detector results (Step 2 of Alg.\ref{selectAlgo})
and later all the consensus results (Step 4 of Alg.\ref{selectAlgo}) to aggregate at both phases of \select.
As such, it is a naive approach that is prone to obtain inferior results in the presence of inaccurate detectors. 

\vspace{-0.1in}
\subsubsection{Selective ensembles}
As motivated earlier in Section \ref{sec:selectmotiv}, carefully selecting which detectors to assemble in Step 2 may help prevent the final ensemble from going astray, provided that
some base detectors may fail to reliably identify the anomalies of interest to a given application. Similarly, pruning away consensus results that may be noisy in Step 4 could help reach a stronger final consensus.  
In anomaly mining, however, it is challenging to identify the components with inferior results given the lack of ground truth to estimate their generalization errors externally. In this section,
we present two orthogonal selection strategies that leverage internal clues across detectors or consensuses and  work in a fully unsupervised fashion: (i) a vertical strategy that exploits correlations among the results, and (ii) a horizontal strategy that uses order statistics to filter out 
far-off results.

\vspace{0.035in}
\noindent
\textbf{Strategy I: Vertical Selection.\;}
Our first approach to selecting the ensemble components is through correlation analysis among the score lists from different methods, based on which we successively enhance the ensemble one list at a time (hence vertical). The work flow of the vertical selection strategy is given in Algorithm \ref{alg:vertical}.

Given a set of anomaly score lists $S$, we first unify the scores by converting them to probability estimates using
{\em Unification} \cite{conf/sdm/KriegelKSZ11}.
Then we average the probability scores across lists to construct a $target$ vector, which we treat as the ``pseudo ground-truth'' (Lines 1-6). 

We initialize the ensemble $E$ with the list $l\in S$ that has the highest weighted Pearson correlation to $target$. In computing the correlation, 
the weights we use for the list elements are equal to $\frac{1}{r}$, where $r$ is the rank of an element in $target$ when sorted in descending order, i.e., the more anomalous elements receive higher weight (Lines 7-11).

Next we sort the remaining lists $S\backslash l$ in descending order by their correlation to the current ``prediction'' of the ensemble, which is defined as the  average probability of lists in the ensemble.
We test whether adding the top list to the ensemble would increase the correlation of the prediction to $target$. If the correlation improves by this addition, we update the ensemble and reorder the remaining lists by their correlation to the updated prediction, otherwise we discard the list.
As such, a list gets either included or discarded at each iteration until all lists are processed (Lines 12-19).

\begin{algorithm}[!t]
\caption{Vertical Selection}
\label{alg:vertical}
\begin{algorithmic}[1]
\REQUIRE $S := \text{set of anomaly score lists}$
\ENSURE $E := \text{ensemble set of selected lists}$ 
\STATE $P := \emptyset$
\STATE $\text{/* convert scores to probability estimates */}$
\FOR{$\text{\bf each }s \in S$}
\STATE $P := P \cup Unification(s)$
\ENDFOR
\STATE $target := avg(P) \text{ \;\;\;  /*target vector*/}$ 
\STATE $r := \text{ranklist after sorting}\; {target}\; \text{in descending order}$
\STATE $E := \emptyset$
\STATE sort $P$ by weighted Pearson ($wP$) correlation  to $target$
\STATE $\text{/* in descending order, weights:}\frac{1}{r}\text{ */}$
\STATE $l := fetchFirst(P)$, \;\; $E := E \cup l$
\WHILE{$P \neq \emptyset$}
\STATE $p := avg(E)$ \;\; /*current prediction of $E$*/ 
\STATE sort $P$ by $wP$ correlation to $p$ \; /*descending order*/
\STATE $l := fetchFirst(P)$
\IF{$\;wP(avg(E\cup l), target) > wP(p, target)$}
\STATE $E := E \cup l \text{ \;\;\;  /*select list*/}$ 
\ENDIF
\ENDWHILE
\RETURN $E$
\end{algorithmic}
\end{algorithm}

\vspace{0.045in}
\noindent
\textbf{Strategy II: Horizontal Selection.}
We are interested in finding time points that are ranked high in a set of accurate rank lists (from either base detectors or consensus methods), ignoring a (small) fraction of inaccurate rank lists. 
Thus, we also present an element-based (hence horizontal) approach for selecting  ensemble components.

To identify the accurate lists, this strategy focuses on the anomalous elements. It
 assumes that the normalized ranks of the anomalies should come from a distribution skewed toward zero. 
Based on this, lists in which the anomalies are not ranked sufficiently high (i.e., have large normalized ranks) are considered to be inaccurate and voted for being discarded.
The work flow of the horizontal selection strategy is given in Algorithm \ref{alg:horizontal}.

Similar to the vertical strategy we first identify a ``pseudo ground truth'', in this case a list of anomalies. 
In particular, we use 
{\em Mixture Modeling} \cite{conf/icdm/GaoT06} to convert each score list in $S$ into a binary list in which outliers are denoted by $1$, and inliers by $0$. 
We then employ majority voting across lists to obtain a final set of target anomalies $O$ (Lines 1-7).

Given that $S$ contains $m$ lists, we construct a normalized rank vector $\mathbf{r}=[r_{(1)}, \ldots, r_{(m)}]$ for each anomaly $o\in O$, such that $r_{(1)} \leq \ldots \leq r_{(m)}$, where $r_{(l)}$ denotes the rank of $o$ in list $l \in S$ normalized by the total number of elements in $l$. 
Following similar ideas to \textit{Robust Rank Aggregation}~\cite{Kolde2012}, we then compute order statistics based on these sorted normalized rank lists to identify the lists that provide statistically large ranks for each anomaly.

Specifically, for each ordered list $l$ in a given $\mathbf{r}$, we compute 
how probable it is to obtain $\hat{r}_{(l)} \leq r_{(l)}$ when the ranks ${\hat{r}}$ are generated by a uniform null distribution. 
We denote the  probability that $\hat{r}_{(l)} \leq r_{(l)}$ by $p_{l,m}(\mathbf{r})$. Under the uniform null model, the probability that ${\hat{r}}(l)$ is smaller or equal to $r_{(l)}$ can be expressed as a binomial probability

\vspace{-0.05in}
$$
p_{l,m}(\mathbf{r}) = \sum_{t=l}^m {m\choose t} r_{(l)}^t (1-r_{(l)})^{m-t},
$$
\vspace{-0.05in}

since at least $l$ normalized rankings drawn uniformly from $[0,1]$ must be in the range $[0,r_{(l)}]$.

\begin{algorithm}[!t]
\caption{Horizontal Selection}
\label{alg:horizontal}
\begin{algorithmic}[1]
\REQUIRE $S := \text{set of anomaly score lists}$
\ENSURE $E := \text{ensemble set of selected lists}$ 
\STATE $M := \emptyset$ ,\; $R := \emptyset$ ,\; $F := \emptyset$ ,\; $E := \emptyset$
\FOR{$\text{\bf each }l \in S$}
\STATE $\text{/* label score lists with 1 (outliers) \& 0 (inliers) */}$
\STATE $class := MixtureModel(l)$ , \;\; $M := M \cup class$
\STATE $R := R \cup ranklist(l)$ 
\ENDFOR
\STATE $O := majorityVoting(M) \text{ \;\;\;  /*target anomalies*/}$ 
\STATE $[S_{sort},pVals] := RobustRankAggregation(R,O)$
\FOR{$\text{\bf each }o \in O$}
\STATE $m_{ind} := \min(pVals(o,:))$ 
\STATE $F := F \cup   S_{sort}(o, (m_{ind}+1):end)$
\ENDFOR
\FOR{$\text{\bf each }l \in S$}
\STATE $count := \text{number of occurrences of}\; l\; \text{in}\; F$
\ENDFOR
\STATE Cluster non-zero $count$s into two clusters, $C_l$ and $C_h$
\STATE $E := S\; \backslash\; \{s\in C_h$\} \; /* discard high-$count$ lists */
\RETURN $E$
\end{algorithmic}
\end{algorithm}
\vspace{-0.8in}

\begin{figure}[h] 
\vspace{-0.1in}
  \centerline{\psfig{figure=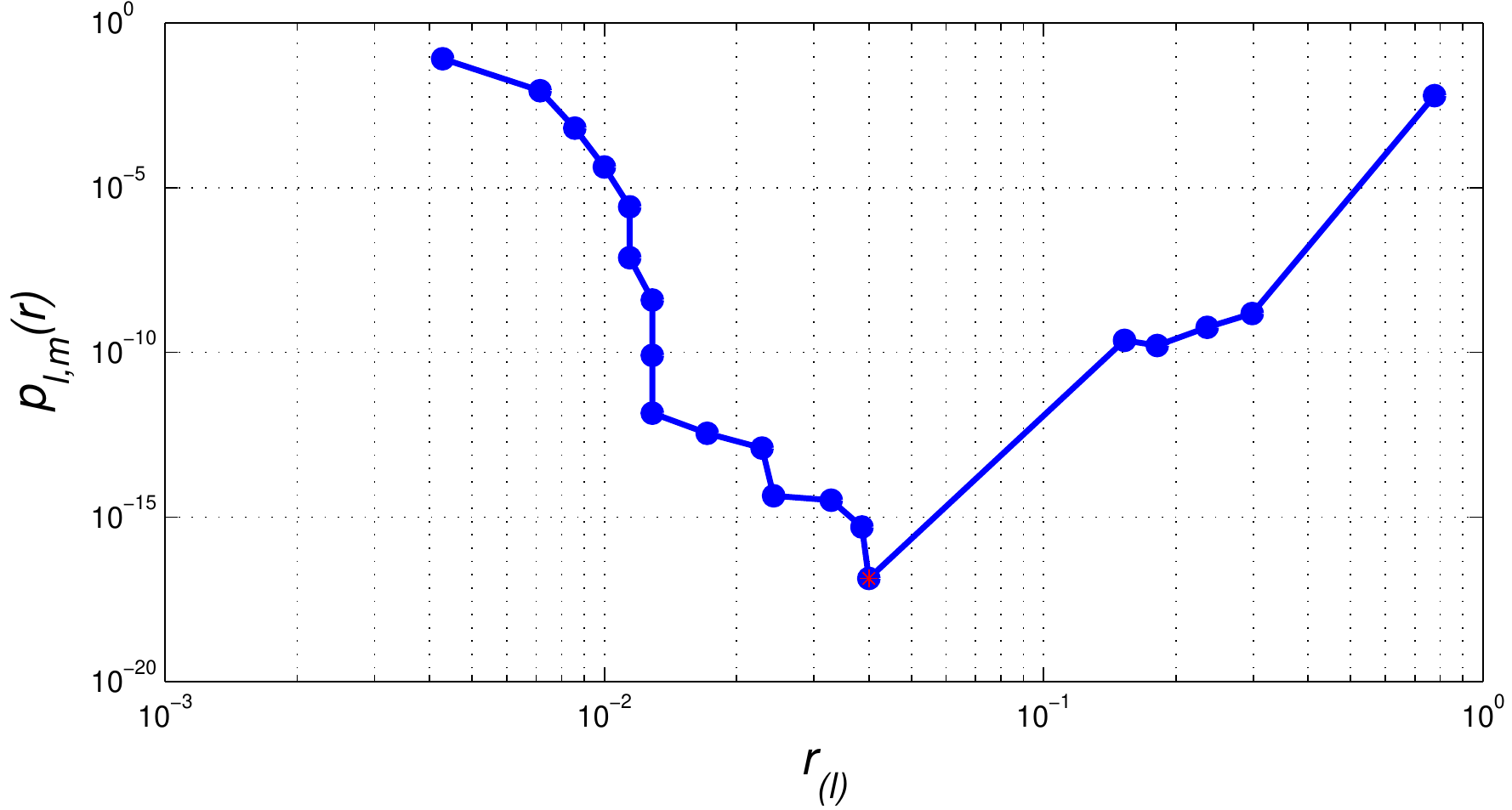,width=2in} }
\vspace{-0.2in}
\caption{\small Normalized rank $r_{(l)}$ vs. probability $p$ that $\hat{r}_{(l)} \leq r_{(l)}$, where $\hat{r}$ are drawn uniformly at random from $[0,1]$.}
\label{fig:orderstat}
\vspace{-0.1in}
\end{figure}

For a sequence of accurate lists that rank the anomalies at the top, and hence that yield low normalized ranks $r_{(l)}$, this probability is expected to drop with the ordering, i.e., for increasing $l \in $ $\{1 \ldots m\}$. 
An example sequence of $p$ probabilities ($y$-axis) are shown in Figure \ref{fig:orderstat} for an anomaly based on $20$ score lists. The lists are sorted by their normalized ranks of the anomaly on the $x$-axis. The figure suggests that the $5$ lists at the end of the ordering are likely inaccurate, as the ranks of the given anomaly in those lists are larger than what is expected based on the ranks in the other lists. 

Based on this intuition, we count the frequency that each list $l$ is ordered \textit{after} the list with $\min_{l=1,\ldots, m} p_{l,m}(\mathbf{r})$ among all the normalized rank lists $\mathbf{r}$ of the target anomalies (Lines 8-15). We then group these counts into two clusters\footnote{We cluster the counts by $k$-means clustering with $k=2$, where the centroids are initialized with the smallest and largest counts, respectively.} and discard the lists in the cluster with the higher average count (Lines 16-17). This way we eliminate the lists with larger counts, but retain the lists that appear inaccurate only a few times which may be a result of the inherent uncertainty or noise in which we construct the target anomaly set.

\subsubsection{Diversity-based ensemble}

In classification, 
two basic conditions for an ensemble to improve over the constituent classifiers are that the base classifiers are (i) accurate (better than random), and (ii) diverse (making uncorrelated  errors) \cite{conf/mcs/Dietterich00,conf/wirn/ValentiniM02}. Achieving better-than-random accuracy in supervised learning is not hard, and several studies have shown that ensembles tend to yield better results when there is a significant diversity among the models \cite{journals/inffus/BrownWHY05,KunchevaWhitaker03}. 

Following on these insights, Schubert {\em et al.} proposed a diversity-based ensemble \cite{conf/sdm/SchubertWZK12}, which is similar to our vertical selection in Alg. \ref{alg:vertical}. The main distinction is the ascending ordering in Lines 9 and 14, which yields a diversity-favored, in contrast to a correlation-favored, selection.\footnote{There are other differences between our vertical selection (Algorithm \ref{alg:vertical}) and the diversity-based ensemble in \cite{conf/sdm/SchubertWZK12}, such as the construction of the pseudo ground truth and the choice of weights in correlation computation.}

Unlike classification ensembles, however, it is not realistic for anomaly ensembles to assume that all the detectors will be reasonably accurate (i.e., better than random), as some may fail to spot the (type of) anomalies in the given data. 
In the existence of inaccurate detectors, 
the diversity-based approach would likely yield inferior results as it is
prone to selecting inaccurate detectors for the sake of diversity.
As we show in our experiments, too much diversity is in fact bound to limit accuracy for event detection ensembles.

\section{Evaluation}
\label{sec:result}

We evaluate our selective ensemble approach on the event detection problem using five real-world datasets, both previously used as well as newly collected by us, including email communications, news corpora, and social media.
For four of these datasets we compiled ground truths for the temporal anomalies, for which we present quantitative results. We use the remaining data for illustrating case studies.

We compare the performance of \select~with vertical selection (\selectv), and horizontal selection (\selecth) to that of individual detectors, 
the full ensemble with no selection (\fulle), and the diversity-based ensemble (\dive) \cite{conf/sdm/SchubertWZK12}.
This makes ours one of the few works that quantitatively compares and contrasts anomaly ensembles at a scale that includes as many datasets with ground truth.

In a nutshell, our results illustrate that ($i$) base detectors do not always all produce accurate results, ($ii$) ensemble
approach alleviates the shortcomings of the inaccurate detectors, ($iii$) a careful selection of ensemble components increases the overall performance, and ($iv$) introducing noisy results decreases overall ensemble accuracy where the diversity-based ensemble is affected the most.

\vspace{-0.1in}
\subsection{Dataset Description}
In the following we describe the five real-world temporal graph datasets we used in this work.
All datasets with ground truth events are made available at {\small \tt \url{http://shebuti.com/SelectiveAnomalyEnsemble/}}.

\noindent
\textbf{Dataset 1: EnronInc.\;}
We use four years (1999--2002) of Enron email communications. In the temporal graphs, the nodes represent email addresses and directed edges depict sent/received relations. Enron email network contains a total of $80,884$ nodes. We analyze the data with daily sample rate skipping the weekends ($700$ time points).
The ground truth captures the major events in the company's history, such as CEO changes, revenue losses, restatements of earnings, etc. 

\vspace{0.015in}
\noindent
\textbf{Dataset 2: RealityMining\;}
Reality Mining is comprised of communication and proximity data of $97$ faculty, student, and staff at MIT recorded continuously via pre-installed software on their mobile devices over $50$ weeks \cite{Eagle08092009}. 
From the raw data we built sequences of weekly temporal graphs for three types of relations; voice calls, short messages, and bluetooth scans. For voice call and short message graphs a directed edge denotes an incoming/outgoing call or message, and for bluetooth graphs an edge depicts physical proximity between two subjects. The ground truth captures semester breaks, exam and sponsor weeks, and holidays.

%

\vspace{0.015in}
\noindent
\textbf{Dataset 3: TwitterSecurity}
We collect tweet samples using the Twitter Streaming API for four months (May 12--Aug 1, 2014).
We filter the tweets containing Department of Homeland Security keywords related to terrorism or domestic security.\footnote{{\footnotesize{\url{http://www.huffingtonpost.com/2012/02/24/homeland-security-manual_n_1299908.html}}}} 
After named entity extraction and resolution (including URLs, hashtags, @ mentions), we build entity-entity co-mention temporal graphs on daily basis ($80$ time ticks).
We compile the ground truth to include major world news of 2014, such as the Turkey mine accident, Boko Haram kidnapping school girls, killings during Yemen raids, etc.

\vspace{0.015in}
\noindent
\textbf{Dataset 4: TwitterWorldCup}
Our Twitter collection also spans the World Cup 2014 season (June 12--July 13). This time, we filter the tweets by popular/official World Cup hashtags, such as 
{\small{{\tt \#worldcup}, {\tt \#fifa}, {\tt \#brazil}}}, etc. 
Similar to TwitterSecurity, we construct entity-entity co-mention temporal graphs on 5 minute sample rate ($8640$ time points).
The ground truth contains the goals, penalties, and injuries in all the matches that involve at least one of the renowned  teams (Brazil, Germany, Argentina, Netherlands, Spain, France).

\vspace{0.015in}
\noindent
\textbf{Dataset 5: NYTNews\;}
This corpus contains all of the published articles in New York Times over 7.5 years (Jan 2000--July 2007) (available from {\small{\url{https://catalog.ldc.upenn.edu/LDC2008T19}}}).
The named entities (people, places, organizations) are hand-annotated by human editors. 
We construct weekly temporal graphs ($390$ time points) in which each node corresponds to a named entity and edges depict co-mention relations in the articles.
The data contains around $320,000$ entities, however no ground truth events.

%

\vspace{-0.1in}
\subsection{Event Detection Performance}


Next we quantitatively evaluate the ensemble methods on detection accuracy. 
The final result output by each ensemble is a rank list, based on which we create the precision-recall (PR) plot for a given ground truth.
We report the area under the PR plot, namely {\em average precision}, as the measure of accuracy.


Table \ref{tab:Enron_10} shows the accuracies for all four ensemble methods on EnronInc., along with the accuracies of the base detectors and consensus methods.
Notice that some detectors yield quite low accuracy (e.g., \asedwout) on this dataset.
Further, \mmmax~consensus provides low accuracy across ensembles no matter which detector results are combined.
\select~ensembles  successfully filter out relatively inferior results and achieve higher accuracy.
We also note that \dive~yields lower performance than all, including \fulle.

To investigate the significance of the selections made by \select~ensembles, we compare them to ensembles that randomly select the same number of components to assemble at each phase. In Table \ref{tab:significance} we report the average and standard deviation of accuracies achieved by 100 such random ensembles, denoted by \rande, and the gain achieved by \selectv~and \selecth~over their respective random ensembles.

\begin{table}[t!]
	\vspace{-0.1in}
	\caption{\small Accuracy of ensembles for EnronInc. (features: weighted in-/out-degree). $*$ depicts selected detector/consensus results. }
	\vspace{-0.25in}
	{\small {
			\begin{center}
				\resizebox{\columnwidth}{!}{%
				\begin{tabular}{l|l|c|c|c|c}
					\hline	
					&       & \fulle            & \dive            & \selectv        &  \selecth  \\
					\hline\hline
					\parbox[t]{-0.1in}{\multirow{10}{*}{\rotatebox[origin=c]{90}{\textit{Base Algorithms}}}}
					& \ebedwin   & \;\;{0.1313} & $*$ & $*$       &  \\
					& \ptsadwin  & \;\;{0.1462} & $*$ &           &  \\
					& \spiritwin & \;\;{0.7032} & $*$ &           & $*$ \\
					& \asedwin   & \;\;{0.5470} & $*$ & $*$       & $*$ \\
					& \maedwin   & \;\;{0.6670} &     &           & $*$ \\
					
					& \ebedwout   & \;\;{0.2846} & $*$ &           &  \\
					& \ptsadwout  & \;\;{0.2118} & $*$ &           &  \\
					& \spiritwout & \;\;{0.4563} & $*$ &           & $*$ \\
					& \asedwout   & \;\;{0.0580} & $*$ &           &  \\
					& \maedwout   & \;\;{0.7328} &           & $*$ & $*$ \\
					\hline 
					\parbox[t]{2mm}{\multirow{7}{*}{\rotatebox[origin=c]{90}{{\em Consensus}}}} 
					& \IR         & $*$\;{0.6829} & $*$\;{0.5660}  & \;\;\;0.6738   & $*$\;{0.8291} \\    
					& \KY	      & $*$\;{0.4086} & $*$\;{0.3703}  & $*$\;{0.6586}  & $*$\;{0.6334} \\    
					& \rra	      & $*$\;{0.6178} & \;\;\;0.4871   & \;\;\;0.5686   & $*$\;{0.6590} \\    
					& \uniavg     & $*$\;{0.5292} & $*$\;{0.5511}  & $*$\;{0.6375}  & $*$\;{0.6207} \\    
					& \unimax     & $*$\;{0.3333} & $*$\;{0.3187}  & \;\;\;0.4314   & $*$\;{0.7353} \\    
					& \mmavg      & $*$\;{0.7513} & $*$\;{0.5726}  & $*$\;{0.7663}  & $*$\;{0.7530} \\    
					& \mmmax      & $*$\;{0.0218} & $*$\;{0.0218}  & \;\;\;0.2108	  & \;\;\;0.0224 \\
					\hline \hline    
					\multicolumn{2}{c} {Final Ensemble} \vline
					& \;\;\;0.7082 &	\;\;\;0.6276 &	\;\;\;0.7125 &	\;\;\;\textbf{0.7920} \\
					\hline    
				\end{tabular}
			}
			\end{center}
		}}
		\vspace{-0.3in}
		\label{tab:Enron_10}
	\end{table}

\begin{figure}[h] 
\vspace{-0.15in}
\center\includegraphics[width=3.25in]{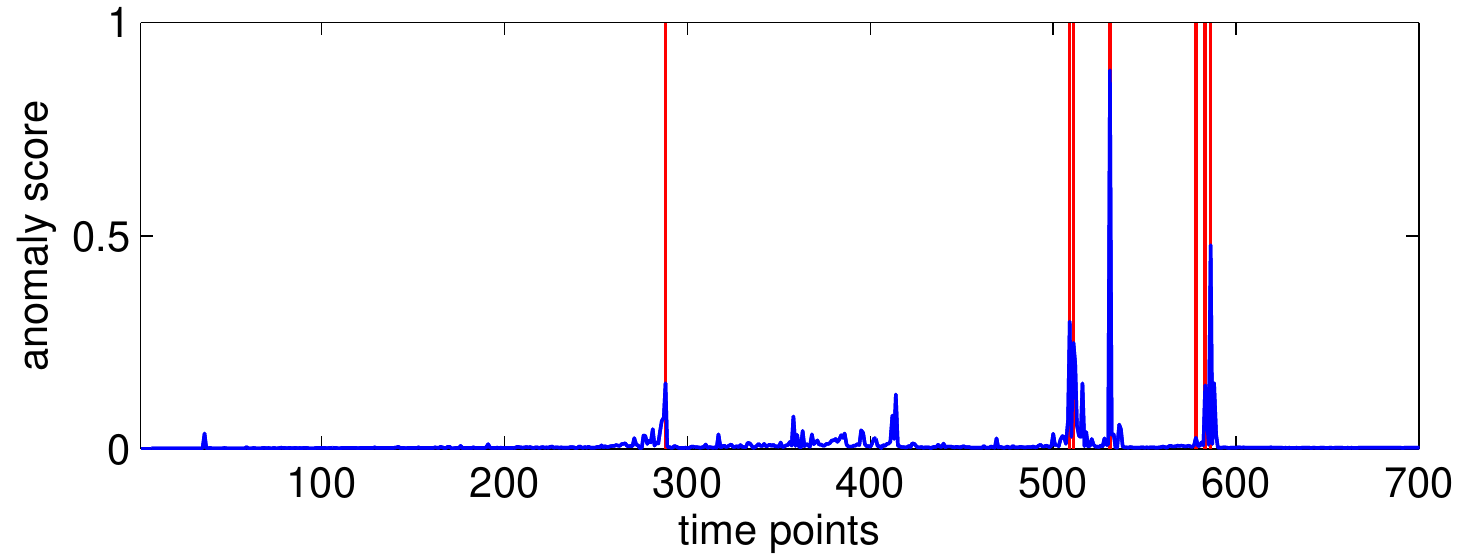}
\vspace{-0.15in}
\caption{\small Anomaly scores of time points by \selecth~on EnronInc. align well with ground truth (vertical red lines).}
\label{fig:enrontimeline}
\end{figure}

 We show the final anomaly scores of the time points provided by \selecth~on EnronInc. for visual analysis in Figure \ref{fig:enrontimeline}.
The figure also depicts the ground truth events by vertical (red) lines, which we note to align well with the time points with high scores.

Table \ref{tab:Enron_10} shows results when we use weighted node in-/out-degree features on the directed Enron graphs to construct the input time series for the base detectors.
As such, the ensembles utilize 10 components in the first phase.
We also build the ensembles using 20 components where we include the unweighted in-/out-degree features.
Table \ref{tab:Enron_20} in Appendix \ref{sec:extraresults} gives all the accuracy results and selections made, a summary of which is provided in Table \ref{tab:significance}.
We notice that the unweighted graph features are less informative and yield lower accuracies across detectors on average. 
This affects the performance of \fulle~and \dive, where the accuracies drop significantly. On the other hand, \select~ensembles are able to achieve comparable accuracies with increased significance under the additional noisy input.

\begin{table}[h!t]
\vspace{-0.1in}
\caption{\small Significance of accuracy results compared to random ensembles with same number of selected components as \select.}
\vspace{-0.1in}
\small{{
		\begin{center} 
			\begin{tabular}{l|l|r} \hline
				  & Accuracy & significance   \\ \hline \hline
			\multicolumn{3}{l}{EnronInc. (10 comp.) (\fulle: 0.7082, \dive: 0.6276)} \\ \hline
				(i) \rande~(3/10, 3/7)  & 0.4804 ($\mu$) &0.1757 ($\sigma$)  \\ 
				\selectv & 0.7125  &$=\mu + 1.3210 \sigma$ \\  \hline
				(ii) \rande~(5/10, 6/7)  & 0.5509 ($\mu$) & 0.1406  ($\sigma$) \\ 
				\selecth & \textbf{0.7920} &$=\mu + 1.7148 \sigma$   \\ \hline \hline

			\multicolumn{3}{l}{EnronInc. (20 comp.) (\fulle: 0.5420, \dive: 0.4697)} \\ \hline
								(i) \rande~(4/20, 2/7)  & 0.4047 ($\mu$) & 0.1732 ($\sigma$)  \\ 
								\selectv & 0.7018  &$=\mu + 1.7154 \sigma$ \\  \hline
								(ii) \rande~(15/20, 6/7)  & 0.5707 ($\mu$)  & 0.0864 ($\sigma$)  \\ 
								\selecth & \textbf{0.7798}  &$=\mu + 2.4201 \sigma$   \\ \hline \hline

			\multicolumn{3}{l}{RM-VoiceCall (10 comp.) (\fulle: 0.7302, \dive: 0.8724)} \\ \hline
				(i) \rande~(2/10, 1/7)  & 0.7370 ($\mu$) & 0.1551 ($\sigma$)  \\ 
								\selectv & 0.8370  &$=\mu + 0.6447 \sigma$ \\ \hline
								(ii) \rande~(8/10, 6/7)  & 0.7653 ($\mu$) & 0.0714 ($\sigma$) \\ 
								\selecth &\textbf{ 0.9045}  &$=\mu + 1.9496 \sigma$   \\ \hline \hline 
								
			\multicolumn{3}{l}{RM-VoiceCall (20 comp.) (\fulle: 0.8011, \dive: 0.8335)} \\ \hline
				(i) \rande~(2/20, 2/7)  & 0.7752 ($\mu$) & 0.1494 ($\sigma$)  \\ 
				\selectv & 0.8847  &$=\mu + 0.7329 \sigma$ \\ \hline
				(ii) \rande~(17/20, 6/7)  & 0.8187 ($\mu$)  & 0.0497 ($\sigma$)  \\ 
				\selecth & \textbf{0.8949}  &$=\mu + 1.5332 \sigma$   \\ \hline \hline 
				
			\multicolumn{3}{l}{RM-Bluetooth (10 comp.) (\fulle: 0.8398, \dive: 0.7735)} \\ \hline
						(i) \rande~(4/10, 1/7)   & 0.8269 ($\mu$) & 0.1129 ($\sigma$)  \\ 
						\selectv & \textbf{0.9193}  &$=\mu + 0.8184 \sigma$ \\ \hline
						(ii) \rande~(8/10, 6/7)   & 0.8410 ($\mu$)  & 0.0322 ($\sigma$)  \\ 
						\selecth & 0.8886  &$=\mu + 1.4783 \sigma$   \\  \hline \hline 
			
			\multicolumn{3}{l}{RM-SMS (10 comp.) (\fulle: 0.9092, \dive: 0.8598)} \\ \hline
						(i) \rande~(4/10, 1/7)  & 0.8328 ($\mu$) & 0.0978 ($\sigma$)  \\ 
						\selectv & \textbf{0.9283}  &$=\mu + 0.9765 \sigma$ \\  \hline
						(ii) \rande~(8/10, 6/7)  & 0.8976 ($\mu$)  & 0.0620 ($\sigma$) \\ 
						\selecth & 0.9217  &$=\mu + 0.3887 \sigma$   \\ \hline \hline 
						
			\multicolumn{3}{l}{RM-SMS (20 comp.) (\fulle: 0.9542, \dive: 0.8749)} \\ \hline			
								(i) \rande~(2/20, 1/7)  & 0.7685 ($\mu$) & 0.1521 ($\sigma$) \\ 
								\selectv & 0.9294  &$=\mu + 1.0579 \sigma$ \\  \hline
								(ii) \rande~(17/20, 5/7)  & 0.9217 ($\mu$)  & 0.0296 ($\sigma$) \\ 
								\selecth & \textbf{0.9621}  &$=\mu + 1.3649 \sigma$   \\ \hline \hline

			\multicolumn{3}{l}{TwitterSecurity (10 comp.) (\fulle: 0.5200, \dive: 0.4800)} \\ \hline
									(i) \rande~(4/10, 1/7)  & 0.5068 ($\mu$) & 0.0755 ($\sigma$) \\ 
									\selectv & 0.5467   &$=\mu + 0.5285 \sigma$ \\  \hline
									(ii) \rande~(9/10, 3/7)  & 0.5198 ($\mu$)   & 0.0538 ($\sigma$) \\ 
									\selecth & \textbf{0.5867}  &$=\mu + 1.2435 \sigma$   \\  \hline \hline

			\end{tabular}
\end{center}
}}
\vspace{-0.45in}
\label{tab:significance}
\end{table}

\begin{figure}[h]
\vspace{-0.125in}
\centering
\begin{tabular}{cc}
\hspace{-0.1in}\includegraphics[width=1.65in,height=1.35in]{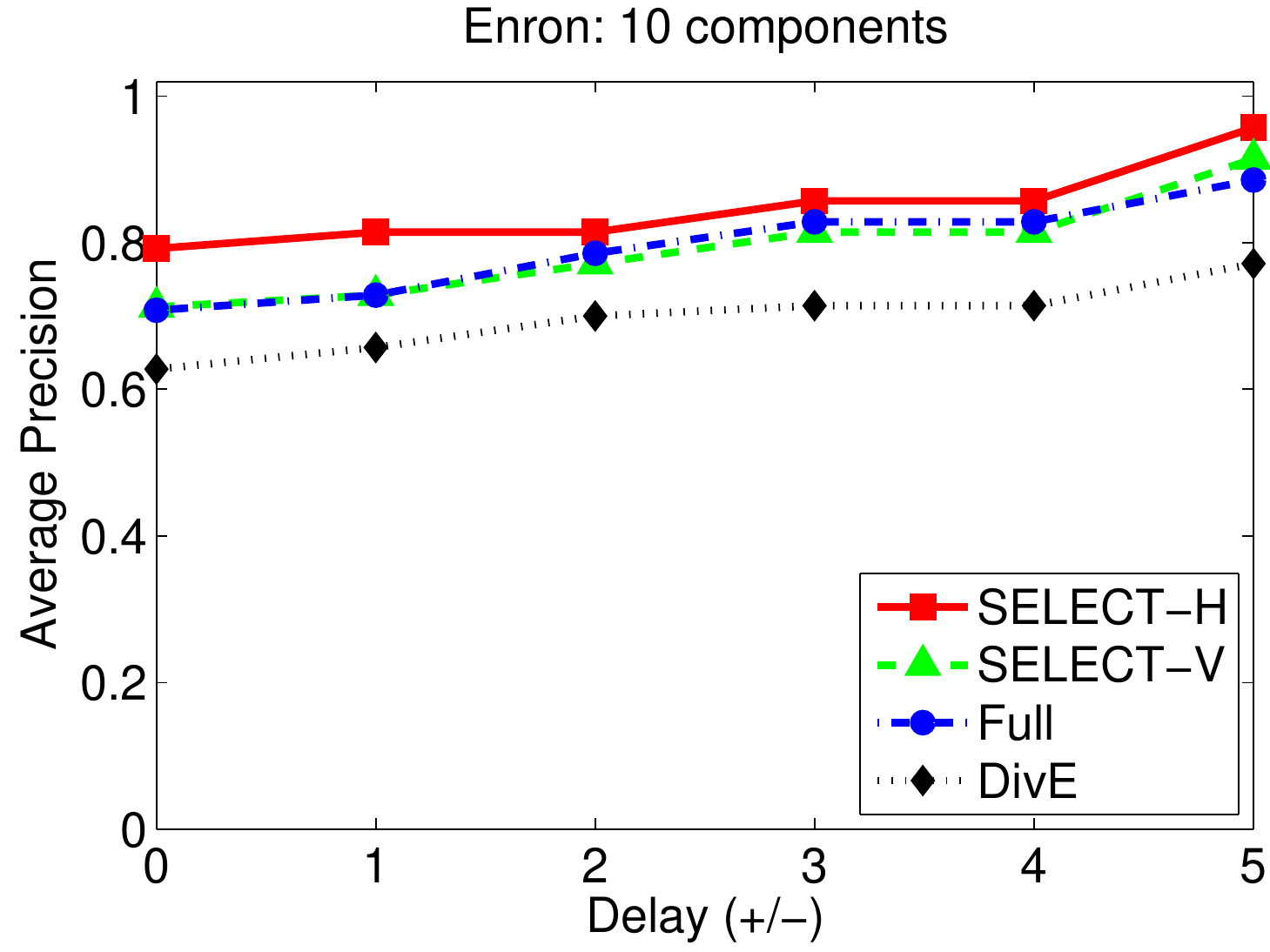} &
\hspace{-0.1in}\includegraphics[width=1.65in,height=1.35in]{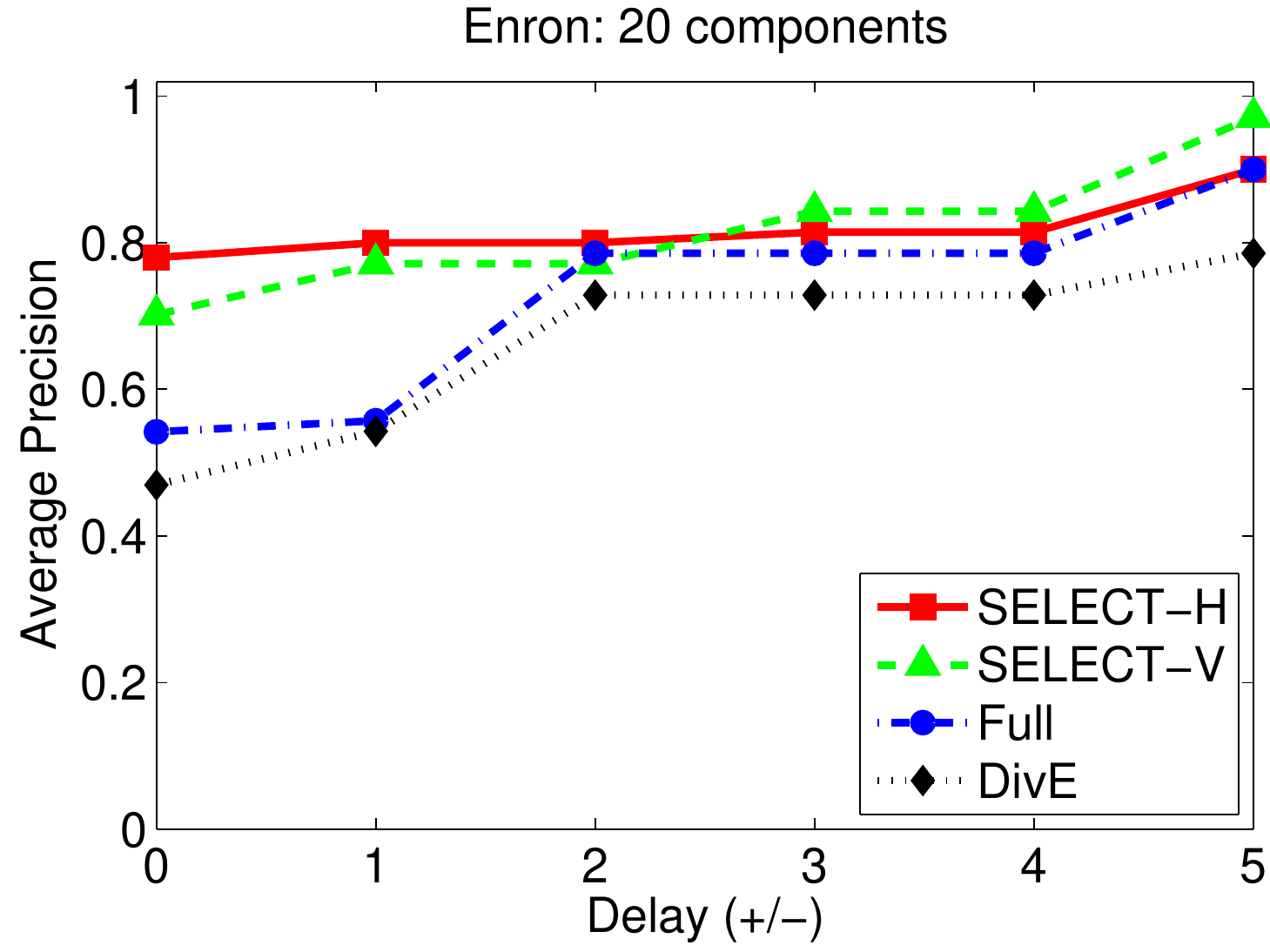} \\
\end{tabular}
\vspace{-0.225in}
\caption{\small EnronInc. average precision vs. detection delay using (left) 10 components and (right) 20 components.} 
\label{fig:Enrondelay}
\vspace{-0.1in}
\end{figure}

Thus far, we used the exact time points of the events to compute precision and recall.
In practice, some time delay in detecting an event is often tolerable.
Therefore, we also compute the detection accuracy when delay is allowed; e.g., for delay $2$, detecting an event that occurred at $t$ within time window
$[t-2,t+2]$ is counted as accurate. Figure \ref{fig:Enrondelay} shows the accuracy for $0$ to $5$ time point delays (days) for EnronInc., where delay $0$ is the same as exact detection.
We notice that \select~ensembles and \fulle~can detect almost all the events within 5 days before or after each event occurs.

\begin{figure*}[!ht]
\centering
\begin{tabular}{ccccc}
\hspace{-0.1in}\includegraphics[width=1.3in,height=1.05in]{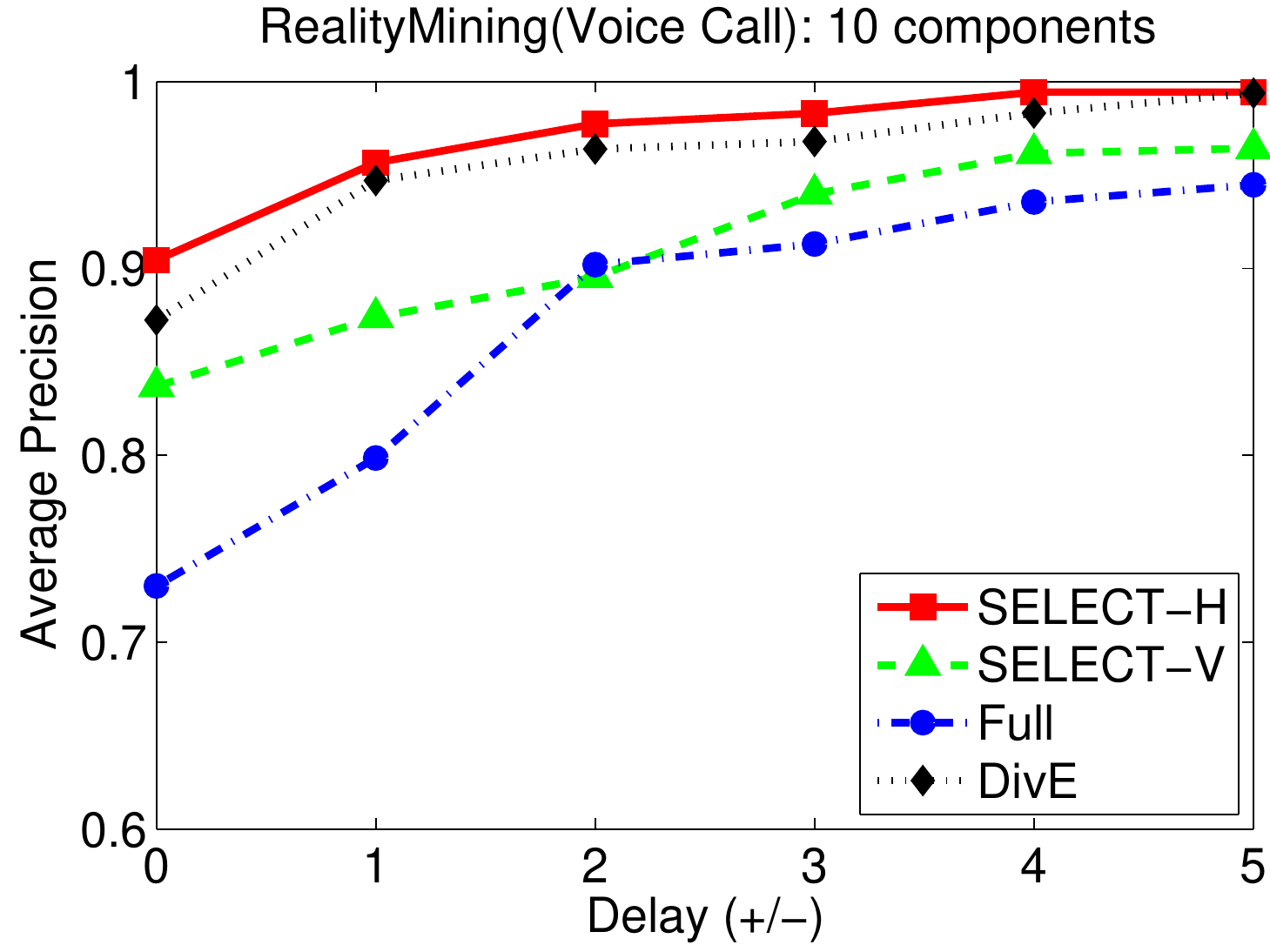} &
\hspace{-0.1in}\includegraphics[width=1.3in,height=1.05in]{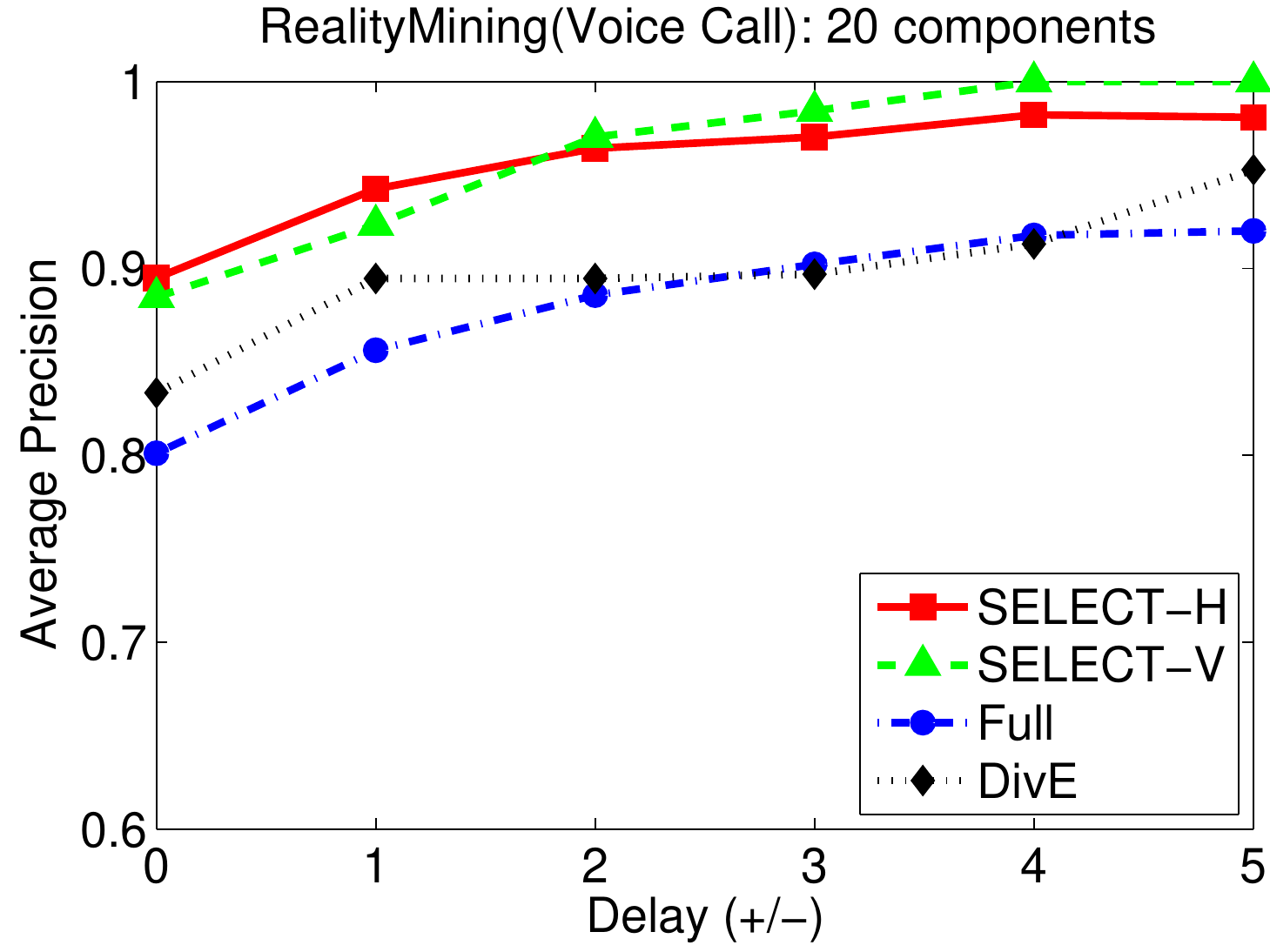} &
\hspace{-0.1in}\includegraphics[width=1.3in,height=1.05in]{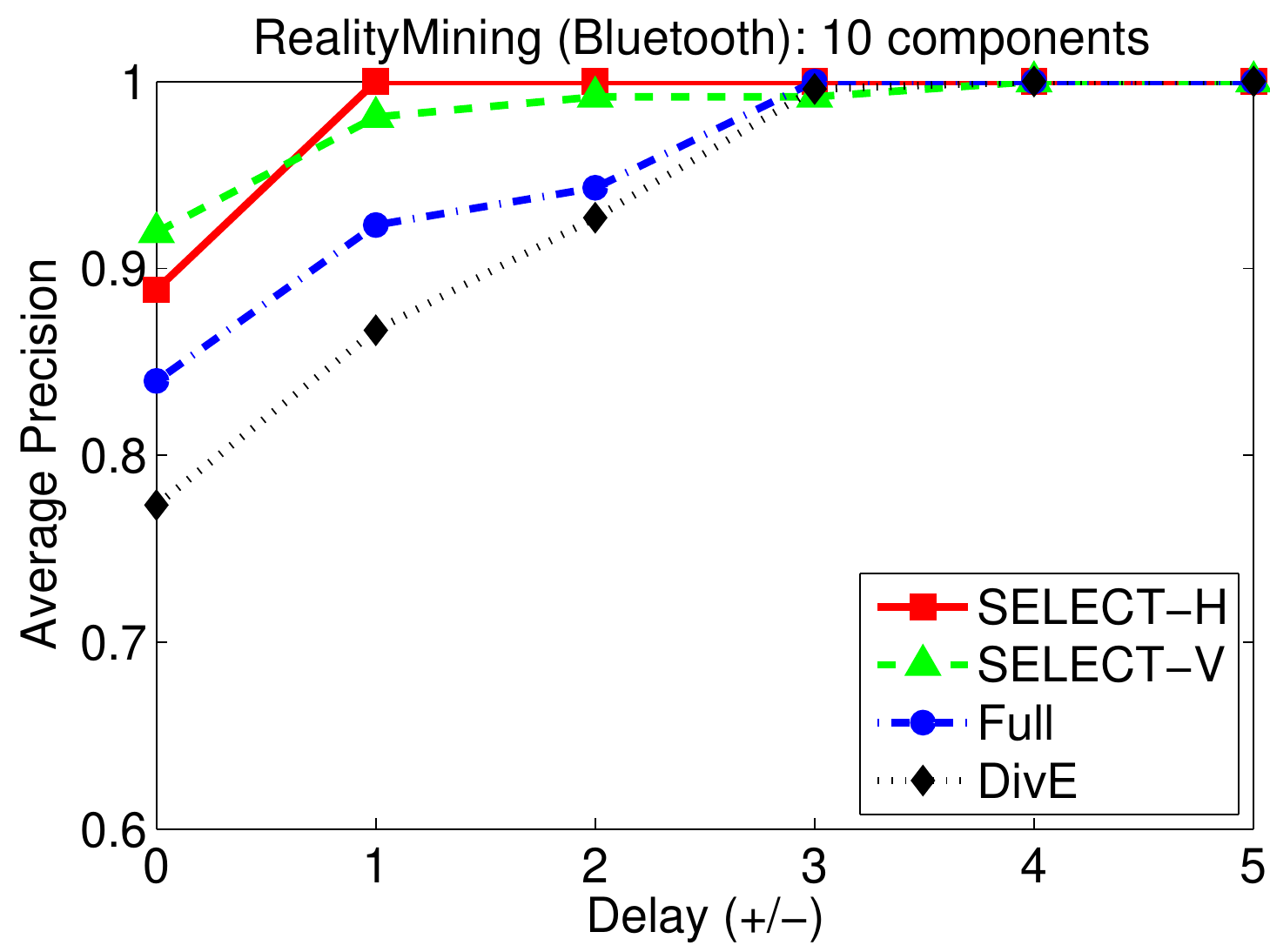} &
\hspace{-0.1in}\includegraphics[width=1.3in,height=1.05in]{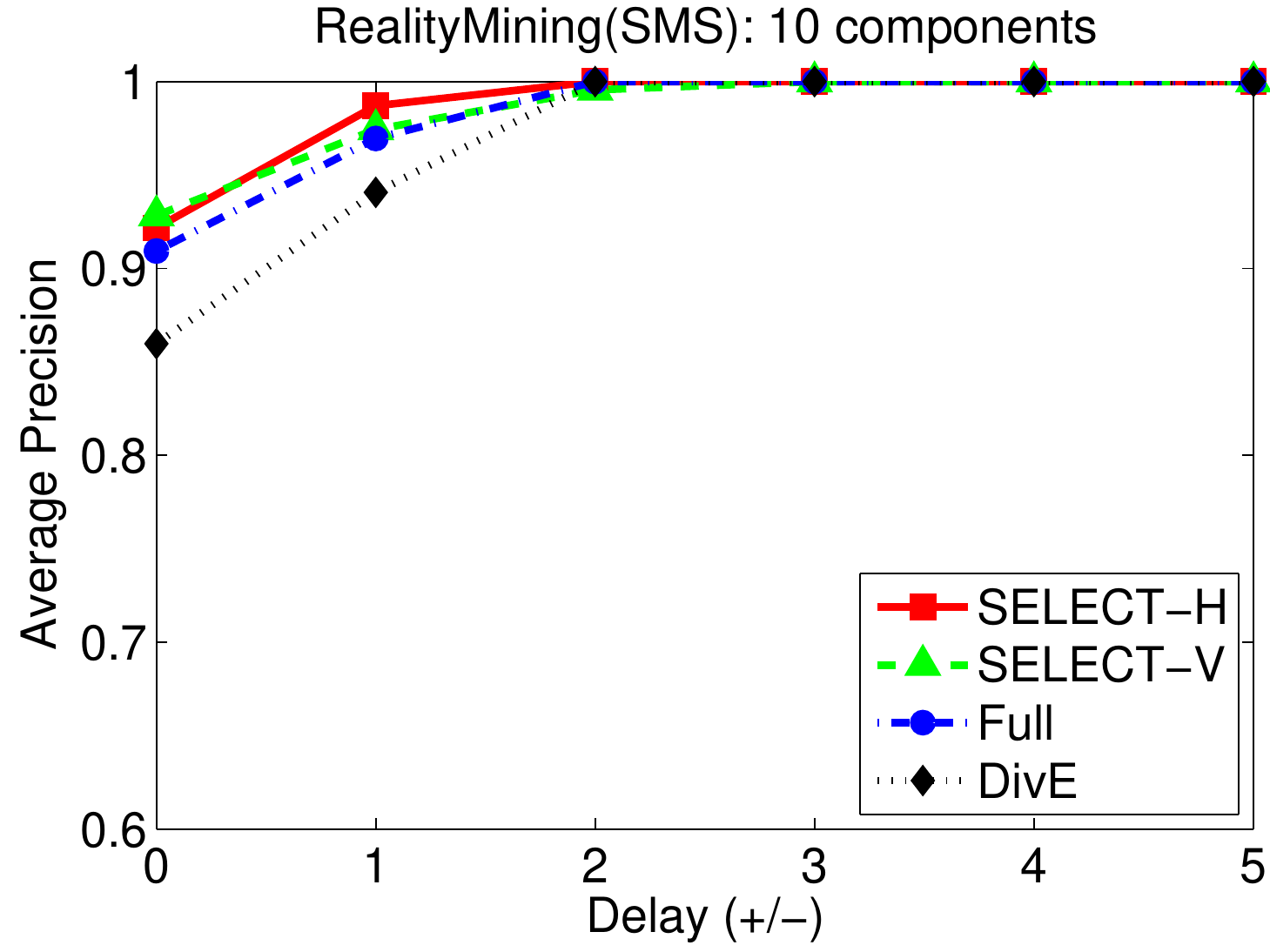} &
\hspace{-0.1in}\includegraphics[width=1.3in,height=1.05in]{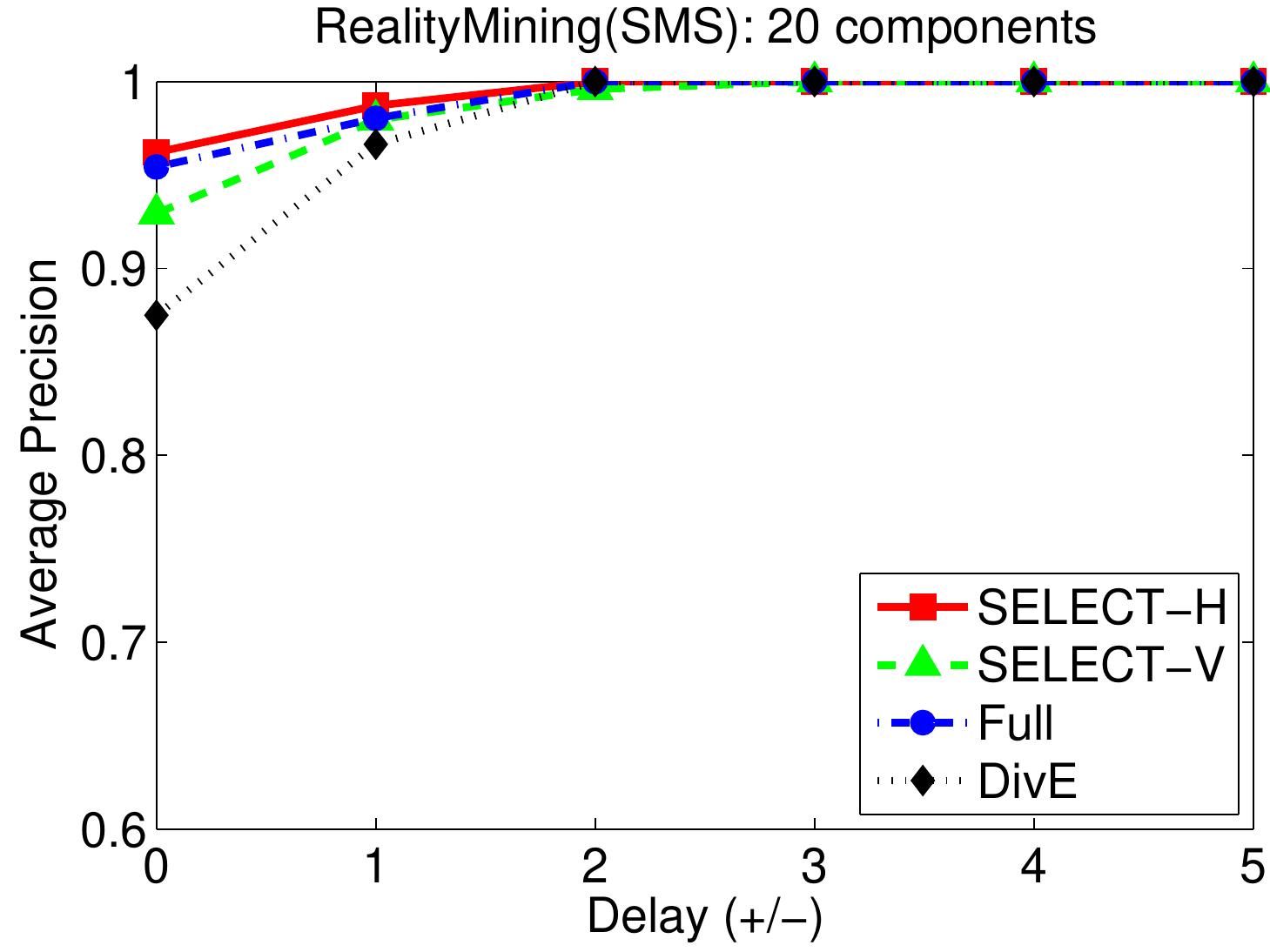}   \\
\end{tabular}
\vspace{-0.2in}
\caption{\small RealityMining average precision vs. detection delay for (left to right) Voice Call (10 comp.), Voice Call (20 comp.), Bluetooth (10 comp.), SMS (10 comp.), and SMS (20 comp.).} 
\label{fig:delayRM}
\vspace{-0.2in}
\end{figure*}

Next we analyze the results for RealityMining. 
Similar to EnronInc., we build the ensembles using both 10 and 20 components for the directed Voice Call and SMS graphs. 
Bluetooth graphs are undirected, as they capture (symmetric) proximity of devices, for which we build
ensembles with 10 components using weighted and unweighted degree features.
All the details on detector and consensus accuracies as well as selections made are given in Appendix \ref{sec:extraresults}
due to space limit (Table \ref{tab:Voice_Call_10} and Table \ref{tab:Voice_Call_20} for Voice Call, Table \ref{tab:Bluetooth_10} for Bluetooth,
Table \ref{tab:SMS_10} and Table \ref{tab:SMS_20} for SMS).
We provide the summary of results in Table \ref{tab:significance}.
We note that \select~ensembles provide superior results to \fulle~and \dive.

Figure \ref{fig:delayRM} illustrates the accuracy-delay plots which show that \select~ensembles for Bluetooth and SMS detect almost all the events within a week before or after they occur, while the changes in Voice Call are relatively less reflective of the changes in the school year calendar.

\begin{figure}[h]
		\vspace{-0.15in}
		\centering
		\begin{tabular}{cc}
		\hspace{-0.1in}\includegraphics[width=1.65in,height=1.35in]{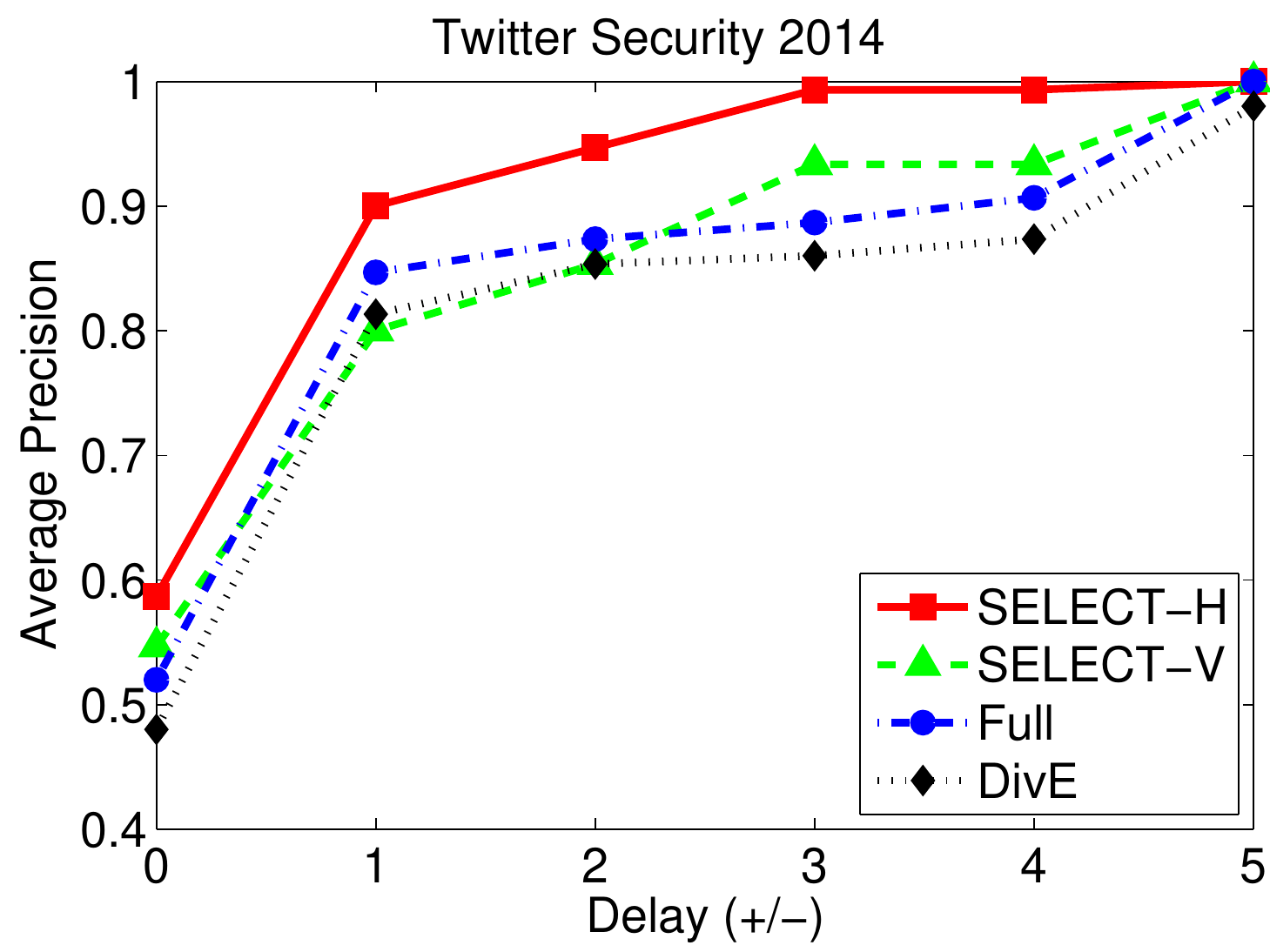} &
		\hspace{-0.1in}\includegraphics[width=1.65in,height=1.35in]{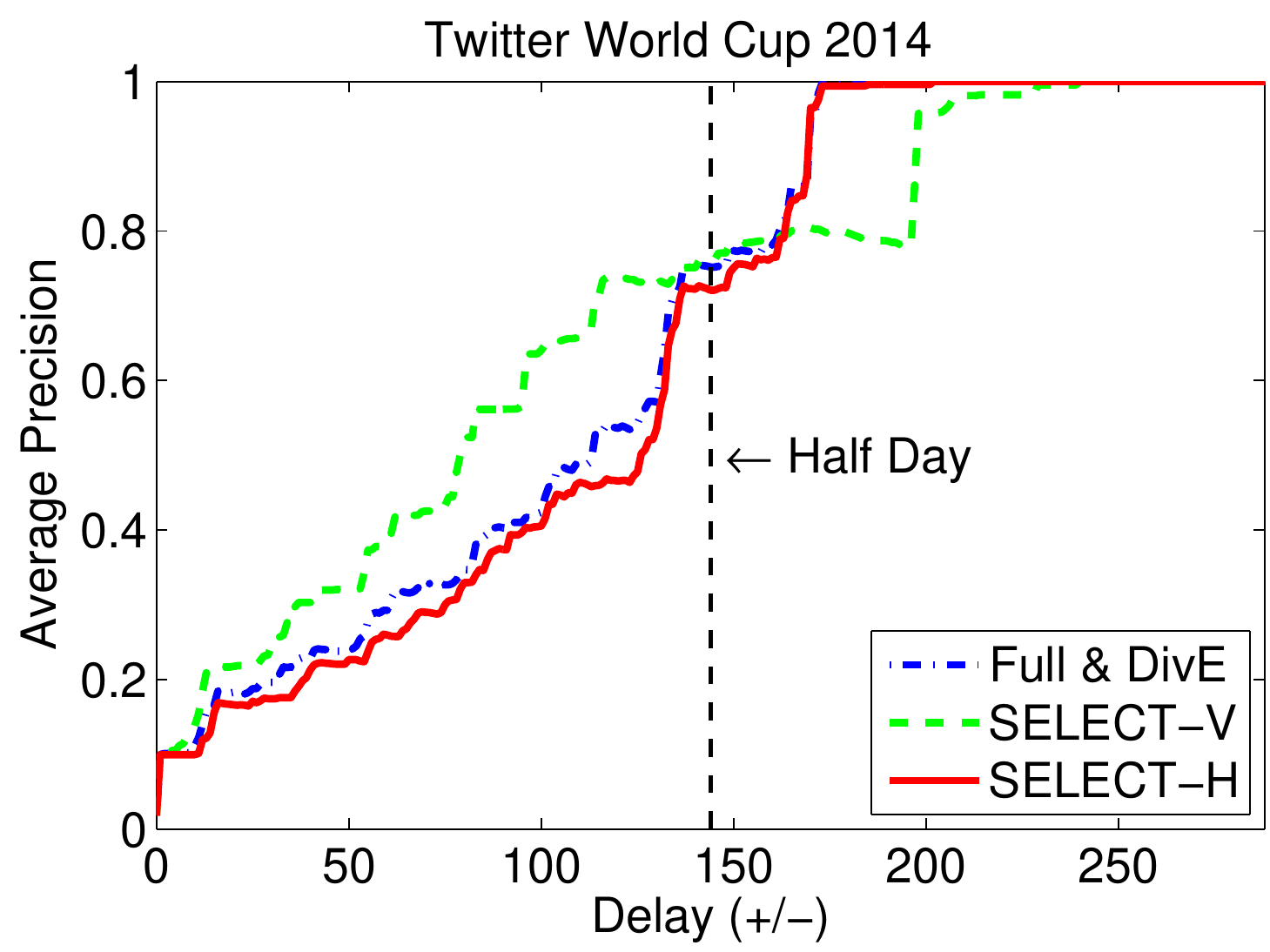} \\
		\end{tabular}
		\vspace{-0.2in}
		\caption{\small Twitter average precision vs. detection delay for (left) Security and (right) WorldCup 2014.} 
		\label{fig:twitterdelay}
		\vspace{-0.25in}
		\end{figure}

		Finally, we study event detection using Twitter. Table \ref{tab:twitter_2014_DSandT_10} in Appendix \ref{sec:extraresults} contains accuracy details for detecting world news on TwitterSecurity, a summary of which is included in 
		Table \ref{tab:significance}. Results are in agreement with prior ones, where \selecth~outperforms the other ensembles.
		This further becomes evident in  Figure \ref{fig:twitterdelay} (left), where \selecth~can detect all the ground truth events within 3 days delay.

		The detection dynamics change when TwitterWorldCup is analyzed. The events in this data such as goals and injuries are quite instantaneous (recall the 4 goals in 6 minutes by Germany against Brazil), where we use a sample rate of 5 minutes. Moreover, such events are likely to be reflected on Twitter with some delay by social media users. As such, it is extremely hard to pinpoint the exact time of the events by the ensembles. As we notice in  Figure \ref{fig:twitterdelay} (right), the initial accuracies at zero delay are quite low. When delay is allowed for up to 288 time points (i.e., one day), the accuracies incline to a reasonable level within half a day delay.
		In addition, all the detector and consensus results seem to contain signals in this case where most of them are selected by the ensembles, hence comparable accuracies. 
		In fact, \dive~selects all of them and performs the same as \fulle.

%

\subsection{Noise Analysis}

\begin{figure*}[!ht]
\centering
\begin{tabular}{ccccc}
\hspace{-0.1in}\includegraphics[width=1.3in,height=1.05in]{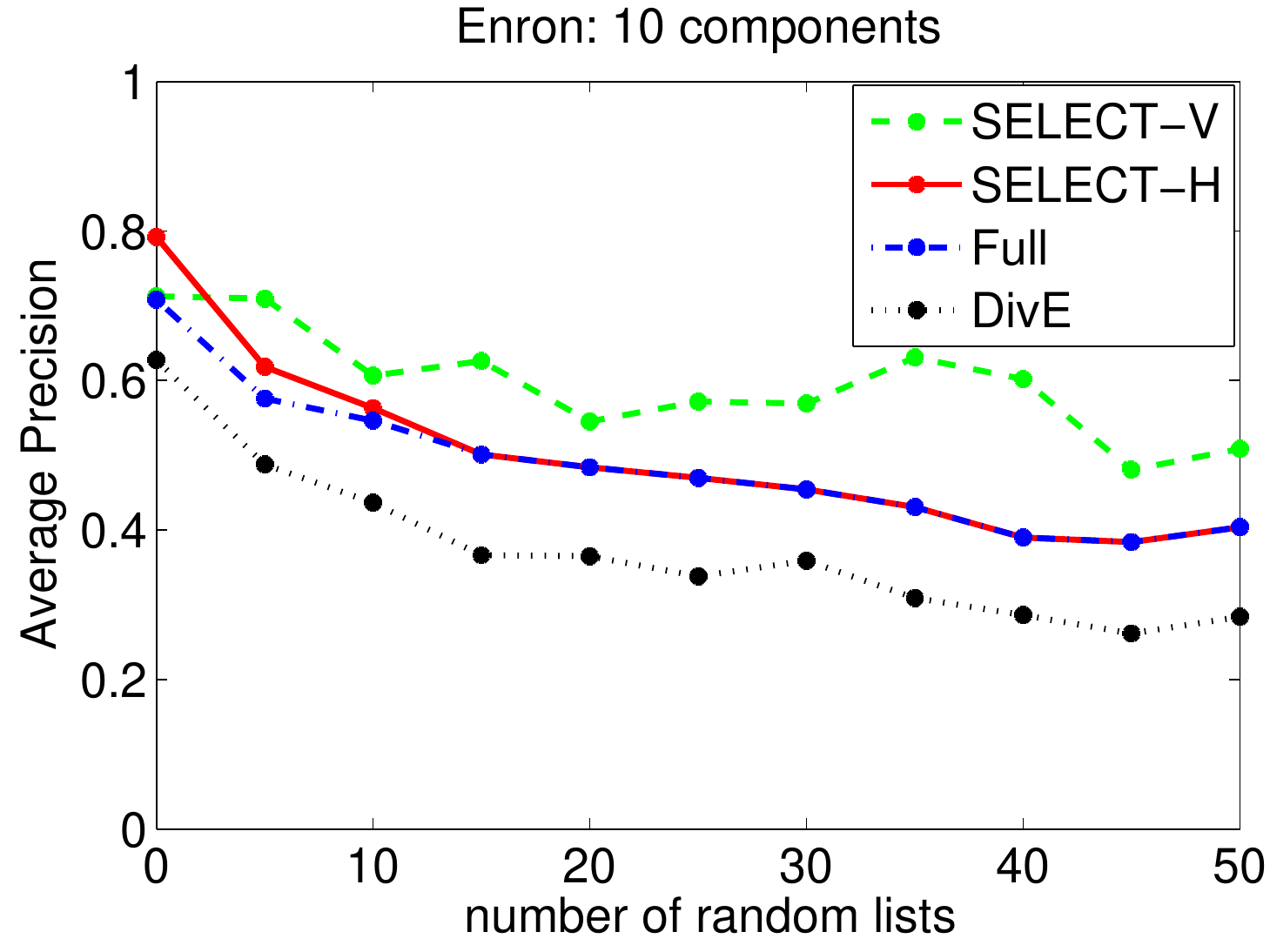} &
\hspace{-0.1in}\includegraphics[width=1.3in,height=1.05in]{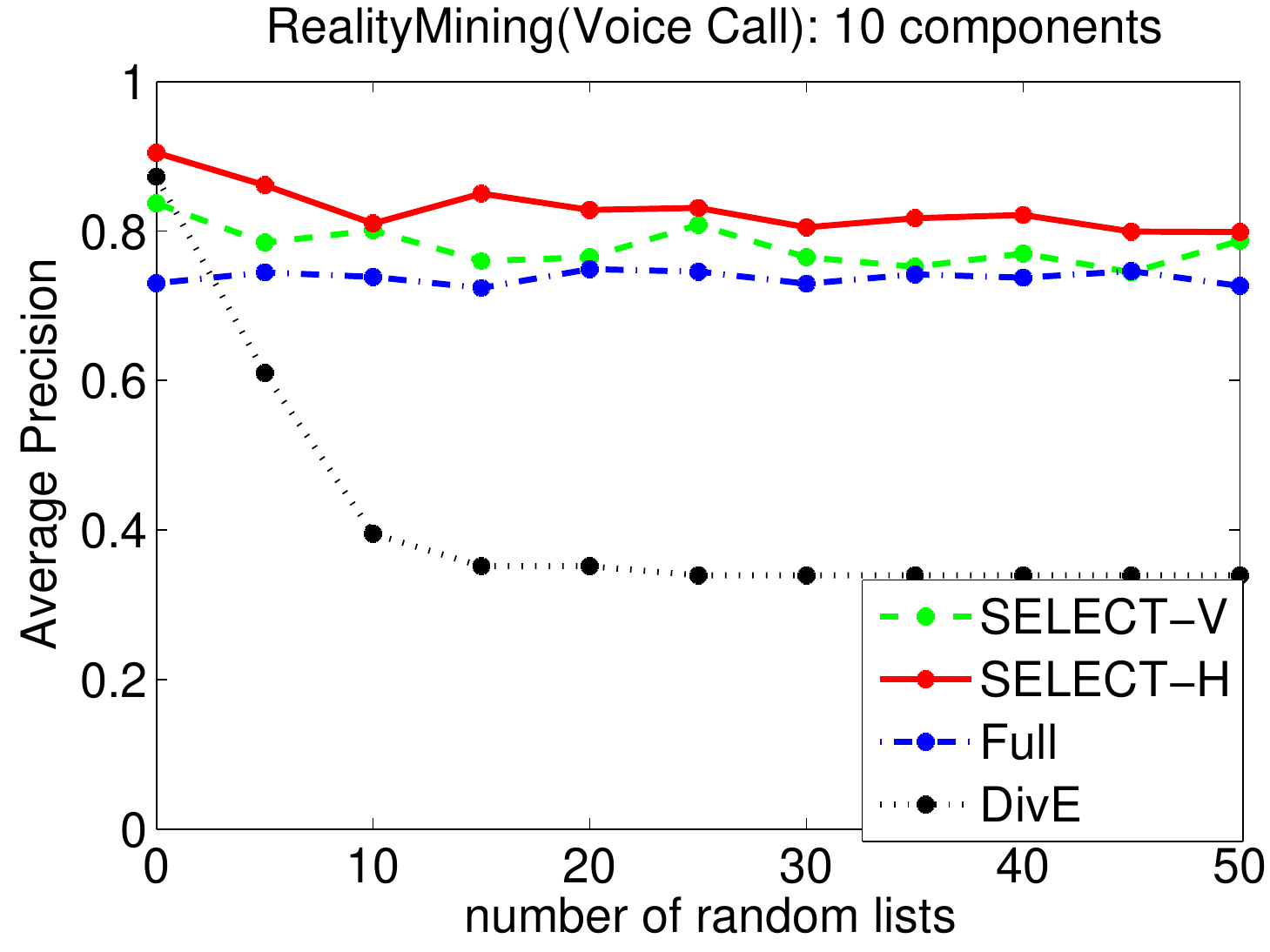} &
\hspace{-0.1in}\includegraphics[width=1.3in,height=1.05in]{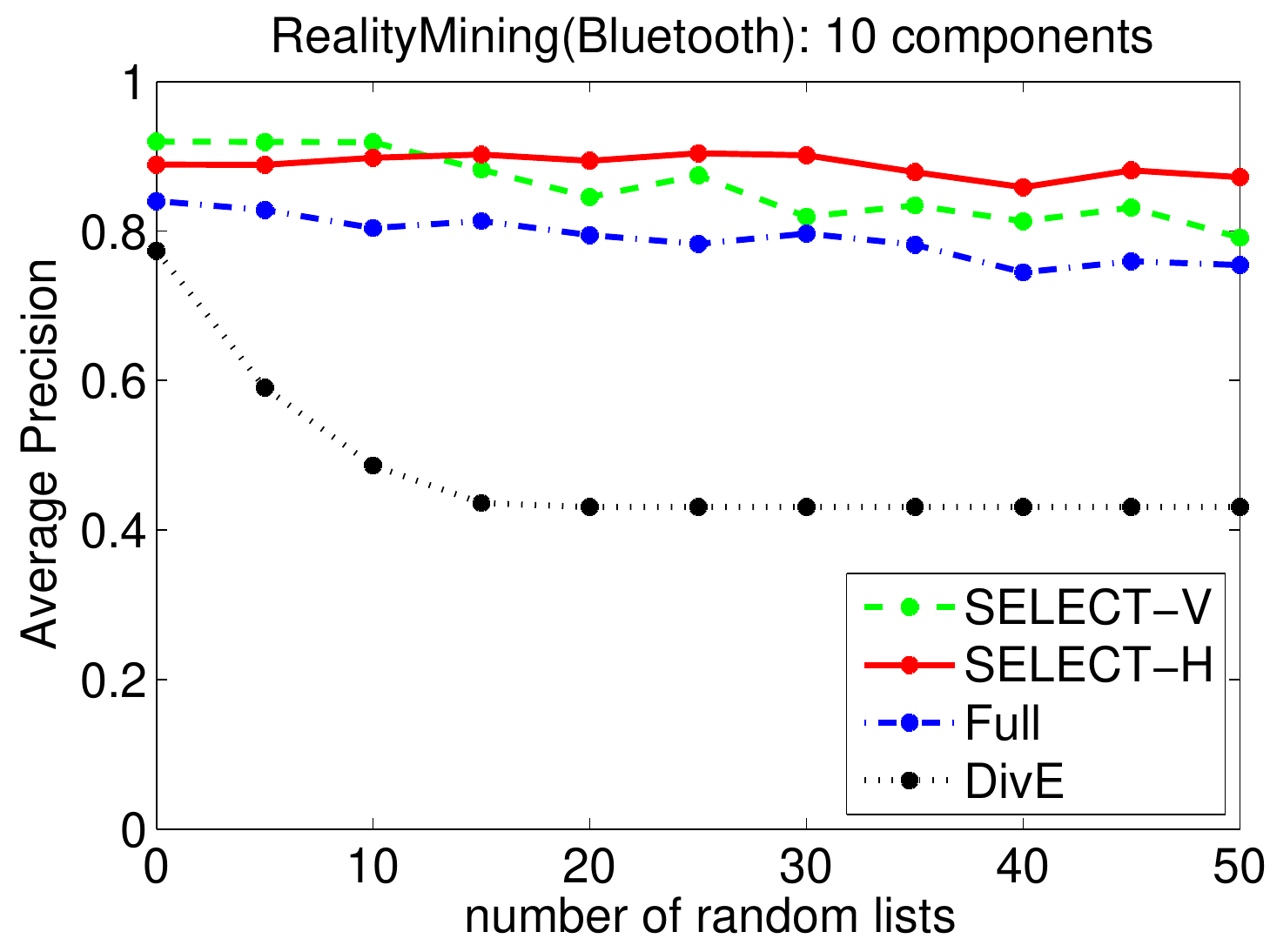} &
\hspace{-0.1in}\includegraphics[width=1.3in,height=1.05in]{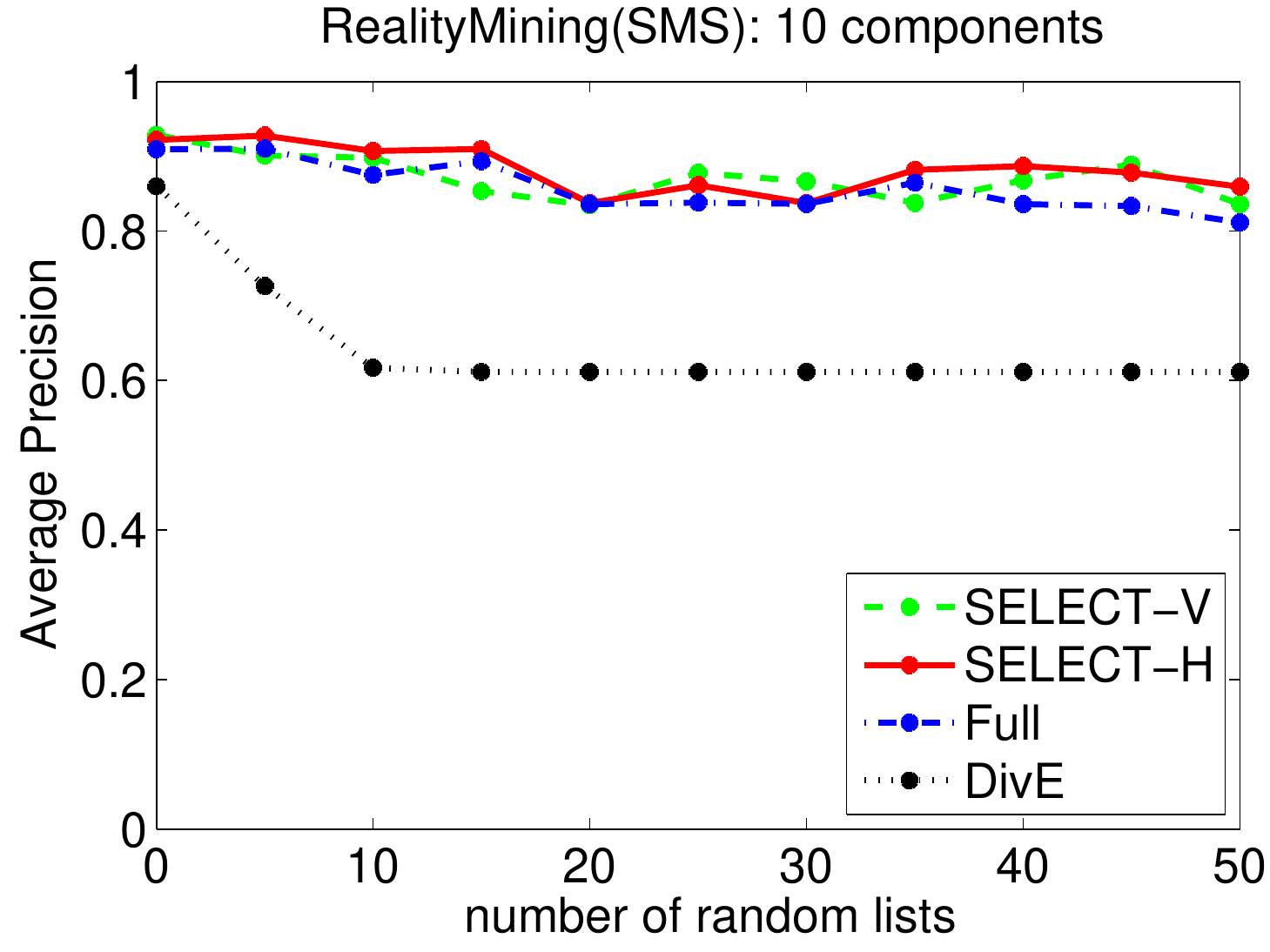} &
\hspace{-0.1in}\includegraphics[width=1.3in,height=1.05in]{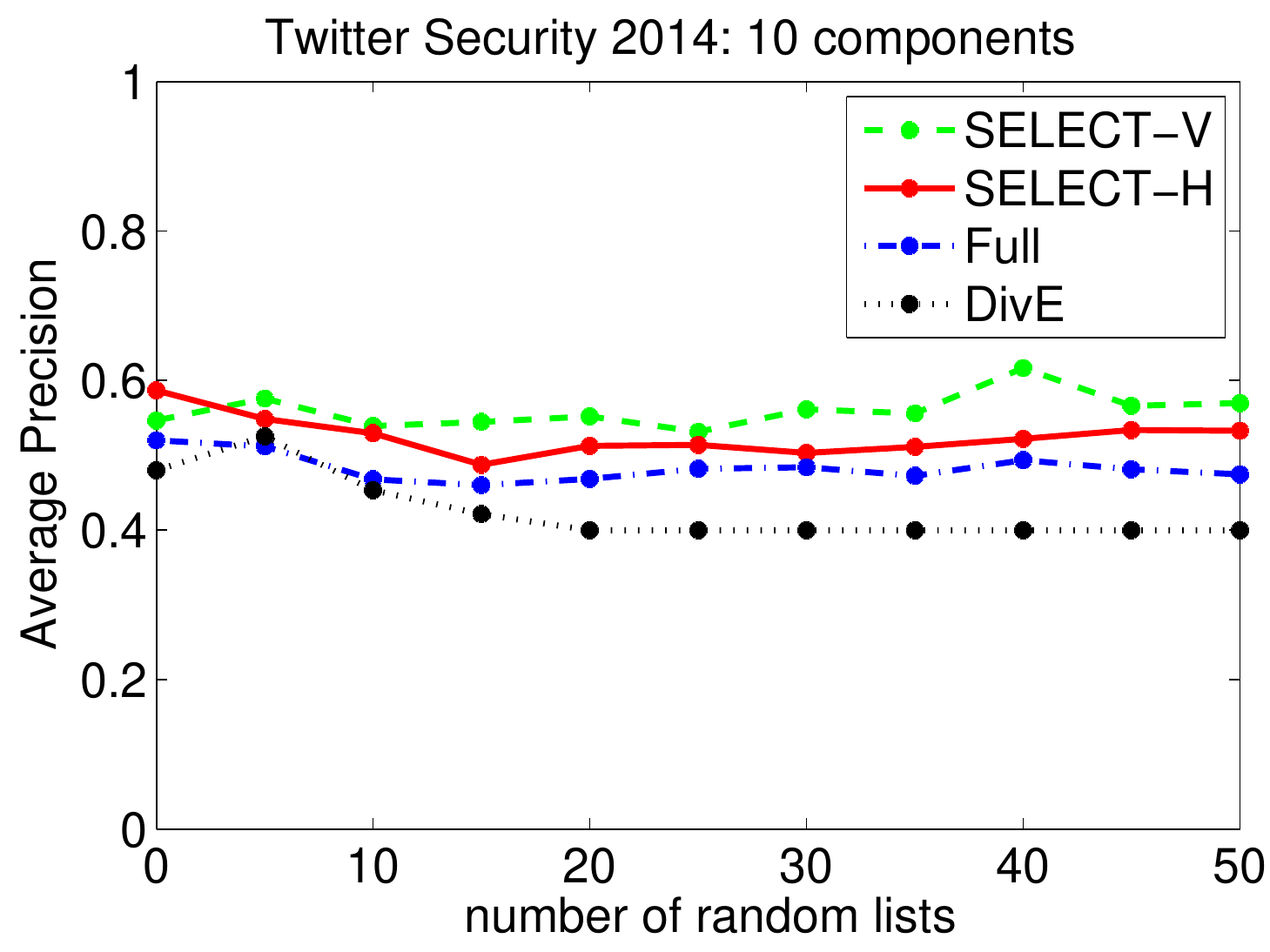}   \\
\end{tabular}
\vspace{-0.2in}
\caption{\small Ensemble accuracies drop when increasing number of random results are added, where decrease is most prominent for {\footnotesize{\dive}}.} 
\label{fig:noise}
\vspace{-0.15in}
\end{figure*}

Provided that selecting which results to combine would especially be beneficial in the presence of inaccurate detectors, we design experiments where we introduce increasing number of noisy results into our ensembles.
In particular, we create noisy results by randomly shuffling the rank lists output by the base detectors and treat them as additional detector results.
Figure \ref{fig:noise} shows accuracies (avg.'ed over 10 independent runs) on all of our datasets for 10 component ensembles (results using 20 components are similar, and provided in Figure \ref{fig:noise2} in Appendix \ref{sec:extraresults}).
We notice that \select~ensembles provide the most stable and effective performance under increasing number of noisy results.
More importantly, these results show that \dive~degenerates quite fast in the presence of noise, i.e., when the assumption that all results are reasonably accurate fails to hold.

	\vspace{-0.1in}
\subsection{Case Studies}

\begin{figure}[!ht]
	\vspace{-0.15in}
	\centerline{\psfig{figure=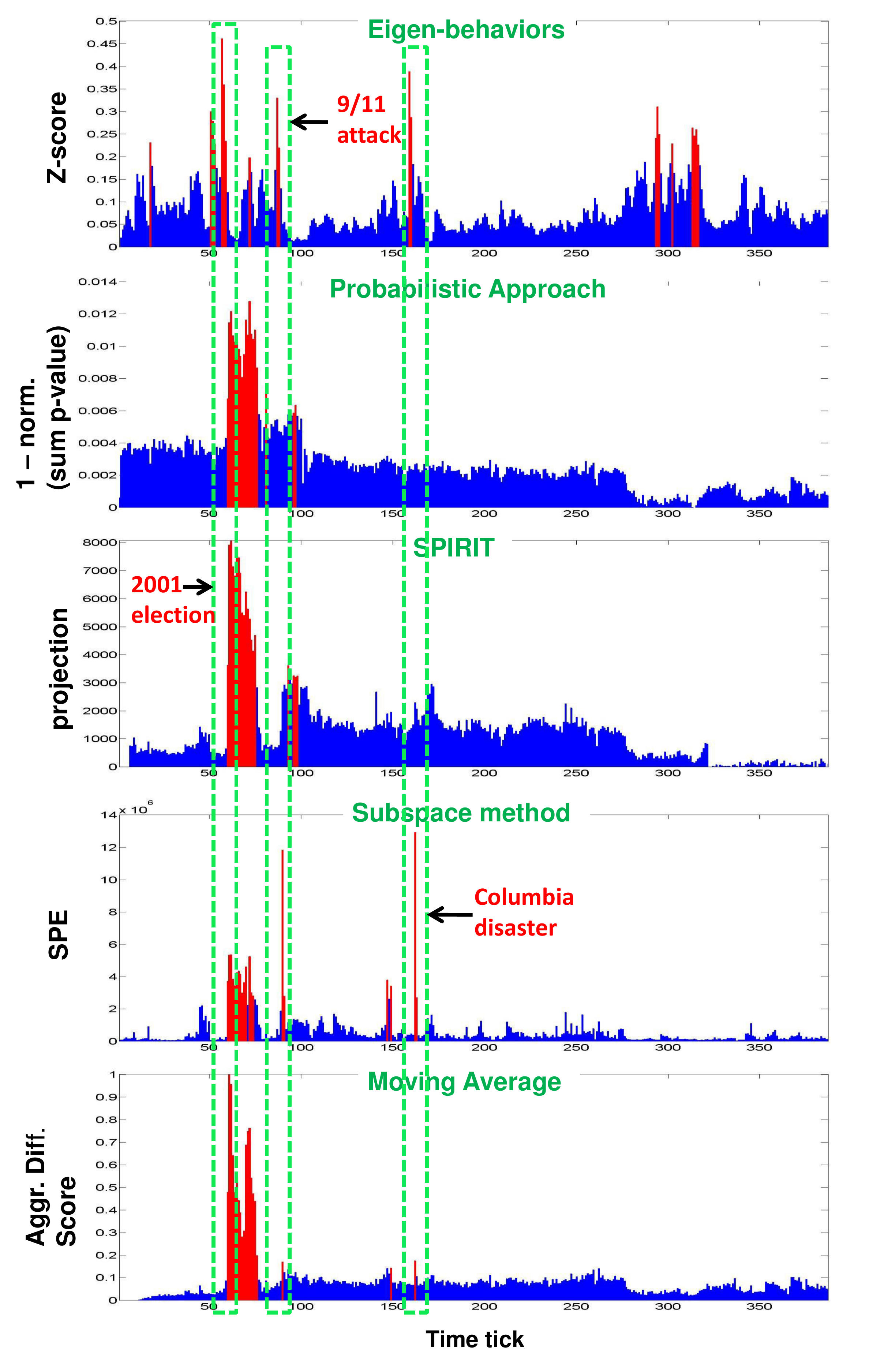,width=\columnwidth, height=4.25in} }
	 \vspace{-0.2in}
	\caption{\small Anomaly scores from five base detectors (rows) for NYT news corpus. Top 3 events by the final ensemble are marked with green boxes. (red bars: top 20 anomalous time points per detector) }
	\vspace{-0.05in}
	\label{fig_NYT_2000_2007_score}
		\vspace{-0.2in}
\end{figure}

In this section we  evaluate our ensemble approach qualitatively using the NYTNews corpus dataset, for which we do not have a compiled list of ground truth events.
Figure \ref{fig_NYT_2000_2007_score} shows the anomaly scores for the 2000-2007 time line, provided by the five base detectors using weighted degree feature (we have demonstrated a similar figure for EnronInc. in Figure \ref{fig:motiv} for additional qualitative analysis).

\begin{figure}[!hb]
	\vspace{-0.05in}
	\centerline{\psfig{figure=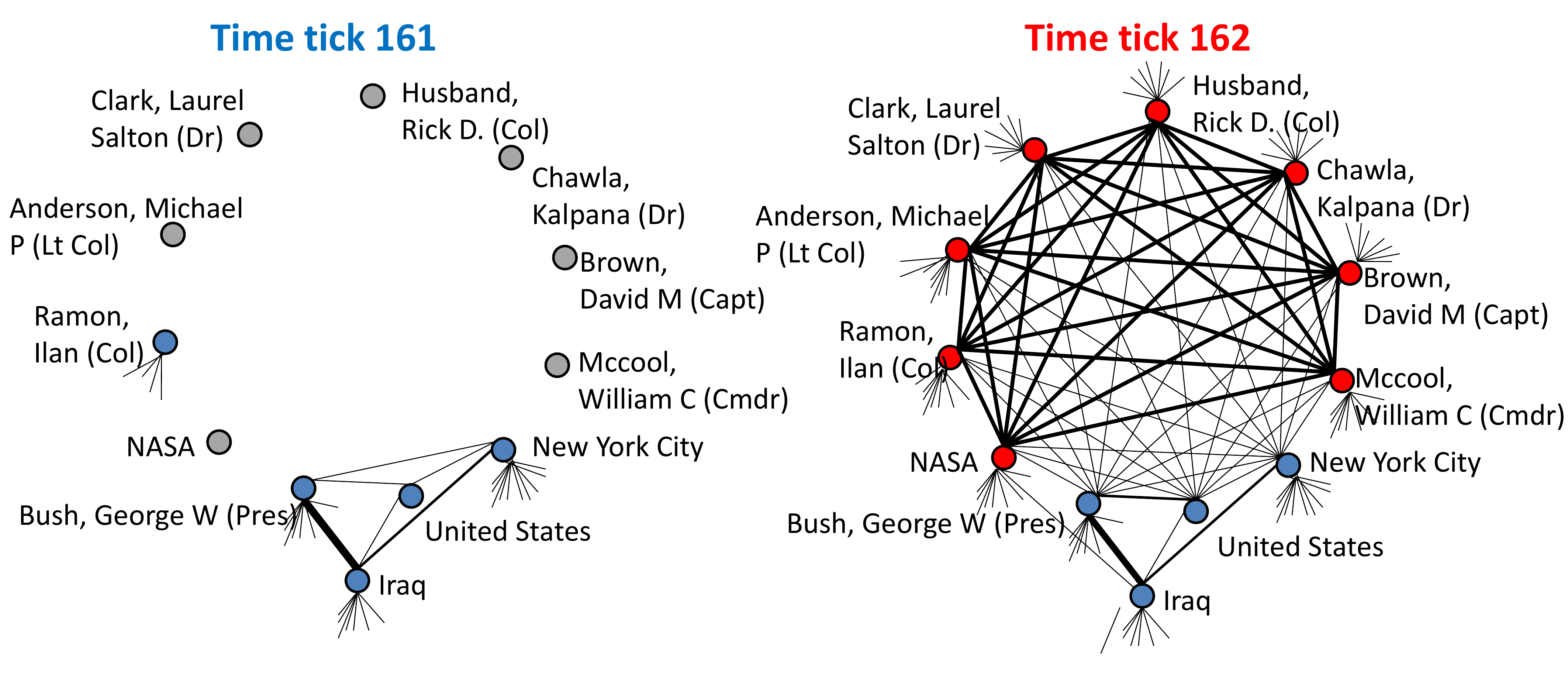,width=0.85\columnwidth} }
	 \vspace{-0.2in}
	\caption{\small During 2003 Columbia disaster a clique of NASA and the seven killed astronauts emerges from time tick 161 to 162.}
	\vspace{-0.05in}
	\label{fig_NYT_columbia_graph}
\end{figure}

Top three events by \selecth~are marked within boxes in the figure, and corresponds to major events such as the 2001 elections, 9/11 WTC attacks, and the 2003 Columbia Space Shuttle disaster.  \selecth~also ranks entities by association to a detected event for attribution. We note that for the Columbia disaster, NASA and the seven astronauts killed in the explosion rank at the top. The visualization of the change in Figure \ref{fig_NYT_columbia_graph} shows that a heavy clique with high degree nodes emerges in the graph structure at the time of the event.


%


\section{Conclusion}
\label{sec:con}

In this work we have proposed \select, a new selective ensemble approach for anomaly mining, and applied it to the event detection problem in temporal graphs.
\select~is a two-phase approach that combines multiple detector results and then multiple consensuses, respectively.
Motivated by our 
earlier observations \cite{RayanaAkoglu14} that inaccurate detectors may deteriorate overall ensemble accuracy, we designed two unsupervised selection strategies, \selectv~and \selecth, which carefully choose which detector/consensus outcomes to assemble.
We compared \select~to \fulle, the ensemble that combines all results, and \dive, an existing ensemble \cite{conf/sdm/SchubertWZK12} that combines diverse, i.e., least correlated results.

Our quantitative evaluation on real-world datasets with ground truth show that building selective ensembles is effective in boosting detection performance.
\selecth~appears to be a better strategy than \selectv, where it either provides the best result (6/8 in Table \ref{tab:significance}) or achieves comparable accuracy when \selectv~is the winner. Selecting results based on diversity turns out to be a poor strategy for anomaly ensembles as \dive~yields even worse results than the \fulle~ensemble (6/8 in Table \ref{tab:significance}). Noise analysis further corroborates the fact that \dive~selects inaccurate/noisy results for the sake of diversity and declines in accuracy much faster than the rest.

Future work will investigate how to go beyond binary selection and estimate weights for the detector/consensus results. We will also apply \select~to the outlier mining problem in multi-dimensional point data and 
continue to enhance it with other detector and consensus methods.

{\scriptsize{
\bibliographystyle{abbrv}
\bibliography{BIB/refsmeas}

\begin{thebibliography}{1}

\bibitem{Akoglu2010army2}
L.~Akoglu and C.~Faloutsos.
\newblock Event detection in time series of mobile communication graphs.
\newblock In {\em 27th Army Science Conference}, 2010.

\bibitem{Cam1998}
A.~Cameron and P.~Trivedi.
\newblock {\em Regression Analysis of Count Data}.
\newblock Cambridge Univ. Press, 1st edition, 1998.

\bibitem{Lakhina04subspace2}
A.~Lakhina, M.~Crovella, and C.~Diot.
\newblock Diagnosing network-wide traffic anomalies.
\newblock In {\em ACM SIGCOMM}, pages 219--230, 2004.

\bibitem{Lam92}
D.~Lambert.
\newblock Zero-inflated poisson regression with an application to defects in
  manufacturing.
\newblock In {\em Technometrics}, pages 1--14, 1992.

\bibitem{Papadimitriou05streamingpattern2}
S.~Papadimitriou, J.~Sun, and C.~Faloutsos.
\newblock Streaming pattern discovery in multiple time-series.
\newblock In {\em VLDB}, pages 697--708, 2005.

\bibitem{Perron1907}
O.~Perron.
\newblock Zur theorie der matrices.
\newblock In {\em Mathematische Annalen}, 1907.

\bibitem{Mich2012}
M.~D. Porter and G.~White.
\newblock Self-exciting hurdle models for terrorist actovity.
\newblock In {\em The Annals of Applied Statistics}, pages 106--124, 2012.

\bibitem{Vuong89}
Q.~H. Vuong.
\newblock Likelihood ratio tests for model selection and non-nested hypotheses.
\newblock In {\em Econometrica}, 1989.

\end{thebibliography}


\begin{thebibliography}{10}

\bibitem{journals/sigkdd/Aggarwal12}
C.~C. Aggarwal.
\newblock Outlier ensembles: position paper.
\newblock {\em SIGKDD Explor. Newsl.}, 14(2):49--58, 2012.

\bibitem{Akoglu2010army}
L.~Akoglu and C.~Faloutsos.
\newblock Event detection in time series of mobile communication graphs.
\newblock In {\em 27th Army Science}, 2010.

\bibitem{DBLP:journals/corr/AkogluTK14}
L.~Akoglu, H.~Tong, and D.~Koutra.
\newblock Graph-based anomaly detection and description: {A} survey.
\newblock {\em DAMI}, 28(4), 2014.

\bibitem{journals/inffus/BrownWHY05}
G.~Brown, J.~L. Wyatt, R.~Harris, and X.~Yao.
\newblock Diversity creation methods: A survey and categorisation.
\newblock {\em Information Fusion}, 6(1):5--20, 2005.

\bibitem{conf/mcs/Dietterich00}
T.~G. Dietterich.
\newblock Ensemble methods in machine learning.
\newblock In {\em Multiple Classifier Systems}, volume 1857 of {\em Lecture
  Notes in Computer Science}, pages 1--15. Springer, 2000.

\bibitem{Dwork2001}
C.~Dwork, R.~Kumar, M.~Naor, and D.~Sivakumar.
\newblock {Rank aggregation methods for the Web}.
\newblock In {\em WWW}, 2001.

\bibitem{Eagle08092009}
N.~Eagle, A.~S. Pentland, and D.~Lazer.
\newblock Inferring friendship network structure by using mobile phone data.
\newblock {\em PNAS}, 2009.

\bibitem{conf/sdm/FernL08}
X.~Z. Fern and W.~Lin.
\newblock Cluster ensemble selection.
\newblock In {\em SDM}, pages 787--797. SIAM, 2008.

\bibitem{conf/pakdd/GaoHZW12}
J.~Gao, W.~Hu, Z.~M. Zhang, and O.~Wu.
\newblock Unsupervised ensemble learning for mining top-n outliers.
\newblock In {\em PAKDD}, 2012.

\bibitem{conf/icdm/GaoT06}
J.~Gao and P.-N. Tan.
\newblock Converting output scores from outlier detection algorithms to
  probability estimates.
\newblock In {\em ICDM}, 2006.

\bibitem{books/crc/aggarwal13/GhoshA13}
J.~Ghosh and A.~Acharya.
\newblock Cluster ensembles: Theory and applications.
\newblock In {\em Data Clustering: Alg. and Appl.} 2013.

\bibitem{journals/inffus/HadjitodorovKT06}
S.~T. Hadjitodorov, L.~I. Kuncheva, and L.~P. Todorova.
\newblock Moderate diversity for better cluster ensembles.
\newblock {\em Information Fusion}, 7(3):264--275, 2006.

\bibitem{journals/pami/HansenS90}
L.~K. Hansen and P.~Salamon.
\newblock Neural network ensembles.
\newblock {\em IEEE Trans. Pattern Anal. Mach. Intell.}, 12(10), 1990.

\bibitem{Kemeny59}
J.~Kemeny.
\newblock Mathematics without numbers.
\newblock In {\em Daedalus}, pages 577--591, 1959.

\bibitem{Kolde2012}
R.~Kolde, S.~Laur, P.~Adler, and J.~Vilo.
\newblock Robust rank aggregation for gene list integration and meta-analysis.
\newblock {\em Bioinformatics}, 28(4):573--580, 2012.

\bibitem{conf/sdm/KriegelKSZ11}
H.-P. Kriegel, P.~Kröger, E.~Schubert, and A.~Zimek.
\newblock Interpreting and unifying outlier scores.
\newblock In {\em SDM}, pages 13--24, 2011.

\bibitem{KunchevaWhitaker03}
L.~Kuncheva and C.~Whitaker.
\newblock Measures of diversity in classifier ensembles and their relationship
  with the ensemble accuracy.
\newblock {\em Machine Learning}, 51:181--207, 2003.

\bibitem{Lakhina04subspace}
A.~Lakhina, M.~Crovella, and C.~Diot.
\newblock Diagnosing network-wide traffic anomalies.
\newblock In {\em SIGCOMM}, pages 219--230, 2004.

\bibitem{conf/kdd/LazarevicK05}
A.~Lazarevic and V.~Kumar.
\newblock Feature bagging for outlier detection.
\newblock In {\em KDD}, pages 157--166. ACM, 2005.

\bibitem{Papadimitriou05streamingpattern}
S.~Papadimitriou, J.~Sun, and C.~Faloutsos.
\newblock Streaming pattern discovery in multiple time-series.
\newblock In {\em VLDB}, 2005.

\bibitem{preisach2007}
C.~Preisach and L.~Schmidt-Thieme.
\newblock Ensembles of relational classifiers.
\newblock {\em Knowl. and Info. Sys.}, 14:249--272, 2007.

\bibitem{RayanaAkoglu14}
S.~Rayana and L.~Akoglu.
\newblock An ensemble approach for event detection in dynamic graphs.
\newblock In {\em KDD ODD$^2$ Workshop}, 2014.

\bibitem{journals/air/Rokach10}
L.~Rokach.
\newblock Ensemble-based classifiers.
\newblock {\em Artif. Intell. Rev.}, 33(1-2):1--39, 2010.

\bibitem{conf/sdm/SchubertWZK12}
E.~Schubert, R.~Wojdanowski, A.~Zimek, and H.-P. Kriegel.
\newblock On evaluation of outlier rankings and outlier scores.
\newblock In {\em SDM}, pages 1047--1058, 2012.

\bibitem{journals/pami/TopchyJP05}
A.~P. Topchy, A.~K. Jain, and W.~F. Punch.
\newblock Clustering ensembles: Models of consensus and weak partitions.
\newblock {\em IEEE Trans. Pattern Anal. Mach. Intell.}, 27(12):1866--1881,
  2005.

\bibitem{conf/wirn/ValentiniM02}
G.~Valentini and F.~Masulli.
\newblock Ensembles of learning machines.
\newblock In {\em WIRN}, 2002.

\bibitem{Zimek13Ensemble}
A.~Zimek, R.~J. Campello, and J.~Sander.
\newblock Ensembles for unsupervised outlier detection: Challenges and research
  questions.
\newblock {\em SIGKDD Explor. Newsl.}, 15(1):11--22, 2013.

\bibitem{conf/kdd/ZimekGCS13}
A.~Zimek, M.~Gaudet, R.~J. G.~B. Campello, and J.~Sander.
\newblock Subsampling for efficient and effective unsupervised outlier
  detection ensembles.
\newblock In {\em KDD}, pages 428--436. ACM, 2013.

\end{thebibliography}
}}

\newpage
\clearpage
\appendix 
\section*{\Large Appendix}
\label{sec:appendix}



\setcounter{section}{0}

\renewcommand\thesection{\Alph{section}}

\vspace{0.05in}

\section{\select~Base Algorithms for Event Detection}
\label{app:base}


\subsection{Eigen Behavior based Event Detection (EBED).} 
The multi-variate time series contain the feature values of each node over time and can be represented as a $n\times t$ data matrix, for $n$ nodes and $t$ time points. 
EBED \citelatex{Akoglu2010army2} defines sliding time windows of length $w$ over the series and computes the principal left singular vector of each $n\times w$ matrix $W$. This vector is the same as the principal eigenvector of $WW^T$ and is always positive due to the Perron-Frobenius theorem \citelatex{Perron1907}.
Each eigenvector $u(t)$ is treated as the ``eigen-behavior''  of the system during time window $t$, the
 entries of which are interpreted as the ``activity'' of each node. 
 
To score the time points, EBED computes the similarity between 
{eigen-behavior} $u(t)$ and a {summary of past eigen-behaviors} $r(t)$, where $r(t)$ is  the arithmetic average of $u(t')$'s for $t'<t$. The anomalousness score of time point $t$ is then $Z = 1-u(t) {\cdot }r(t) \in [0,1]$, where high value of $Z$ indicates a change point. 
For each anomalous time point $\bar{t}$, EBED performs attribution by computing the relative change $\frac{|u_{i}(\bar{t})-r_{i}(\bar{t})|}{u_{i}(\bar{t})}$ of each node $i$ at $\bar{t}$. The higher the relative change, the more anomalous the node is.

\noindent
\subsection{Probabilistic Time Series Anomaly Detection (PTSAD).\;}
A common approach to time series anomaly detection is to probabilistically model a given series and detect anomalous time points based on their likelihood under the model.
PTSAD models each series with four different parametric models and performs model selection to identify the best fit for each series. Our first model is the \textit{Poisson}, which is used often for fitting count data.
However, Poisson is not sufficient for sparse series with many zeros. Since real-world data is frequently characterized by over-dispersion and excess number of zeros,  
we employ a second model called {\em Zero-Inflated Poisson} (ZIP)~\citelatex{Lam92} to account for data sparsity.

We further look for simpler models which fit data with many zeros and employ the {\em Hurdle models}~\citelatex{Mich2012}. Rather than using a single but complex distribution, Hurdle models assume that the data is generated by two simple, separate processes;  (i) the hurdle and (ii) the count processes.  The hurdle process determines whether there exists activity at a given time point and in case of activity the count process determines the actual (positive) counts. For the hurdle process, we employ two different models. First is the independent {\em Bernoulli} and the second is the first order {\em Markov} model which
better captures the dependencies, where an activity influences the probability of subsequent activities.
For the count process, we use the {\em Zero-Truncated Poisson} (ZTP)~\citelatex{Cam1998}.

Overall we model each time series with four different models: Poisson, ZIP, Bernoulli+ZTP and Markov+ZTP. 
We then employ Vuong's likelihood ratio test~\citelatex{Vuong89} to select the best model for individual series.
Note that the best-fit model for each series may be different.

To score the time points, we perform a single-sided test to compute a $p$-value for each value $x$ in a given series; i.e., $P(X \ge x) = 1 - cdf_{H}(x) + pdf_{H}(x)$, where $H$ is the best-fit model for the series. The lower the $p$-value, the more anomalous the time point is.
We then aggregate all the $p$-values from all the series per time point by taking the normalized sum of the $p$-values and inverting them to obtain scores $\in [0,1]$ (s.t. higher is more anomalous).
For each anomalous time point $\bar{t}$, attribution is done by sorting the nodes (i.e., the series) based on their $p$-values at $\bar{t}$.

\subsection{Streaming Pattern DIscoveRy in multIple Time-Series (SPIRIT).\;}
{SPIRIT}~\citelatex{Papadimitriou05streamingpattern2} can incrementally capture correlations, discover trends, and dynamically detect change points in multi-variate time series. The main idea is to represent the underlying trends of a large number of numerical streams with a few hidden variables, where the hidden variables are the \textit{projections} of the observed streams onto the principal direction vectors (eigenvectors). These discovered trends are exploited for detecting change points in the series. 

The algorithm starts with a specific number of hidden variables that capture the main trends of the data. Whenever the main trends change, new hidden variables are introduced or several of existing ones are discarded to capture the change. SPIRIT can further quantify the change in the individual time series for attribution through their \textit{participation weights}, which are the entries in the principal direction vectors. For further details on the algorithm, we refer the reader to the original paper by Papadimitriou \textit{et al.} \citelatex{Papadimitriou05streamingpattern2}.

\subsection{Anomalous Subspace based Event Detection (ASED).}
ASED~\citelatex{Lakhina04subspace2} is based on the separation of high-dimensional space occupied by the time series into two disjoint subspaces, the normal and the anomalous subspaces. {Principal Component Analysis} is used to separate the high-dimensional space, where the major principal components capture the most variance of the data and hence, construct the normal subspace and the minor principal components capture the anomalous subspace. The projection of the time series data onto these two subspaces reflect the normal and anomalous behavior. To score the time points, ASED uses the \textit{squared prediction error} (SPE) of the residuals in the anomalous subspace. The residual values associated with individual series at the anomalous time points are used to measure the anomalousness of nodes for attribution.
For the specifics of the algorithm, we refer to the original paper by Lakhina \textit{et al.} ~\citelatex{Lakhina04subspace2}.

\subsection{Moving Average based Event Detection (MAED).}
MAED is a simple approach that calculates the moving average $\mu_t$ and the moving standard deviation $\sigma_t$ of each time series corresponding to each node by extending the time window one point at a time. If the value at a specific time point is more than three moving standard deviations away from the mean, then the point is considered as anomalous and assigned a non-zero score. The anomalousness score is the difference between the original value and ($\mu_t+3\sigma_t$) at $t$. To score the time points collectively, MAED aggregates their scores across all the series. 
For each anomalous time point $\bar{t}$, attribution is done by sorting the nodes (i.e., the series) based on the individual scores they assign to $\bar{t}$.

\section{Additional Evaluation Results}
\label{sec:extraresults}

\begin{table}[H]
\caption{\small Accuracy of ensembles for EnronInc. (directed) (20 components)
(features: weighted in-/out-degree and unweighted in-/out-degree).\;\;
$*$ depicts selected detector/consensus results.}
\vspace{-0.25in}
{\small {
		\begin{center}
			\resizebox{\columnwidth}{!}{%
			\begin{tabular}{c|l|c|c|c|c}
				\hline	
				&      & \fulle            & \dive            & \selectv       & \selecth       \\
				\hline\hline
				\parbox[t]{2mm}{\multirow{20}{*}{\rotatebox[origin=c]{90}{\textit{Base Algorithms}}}}
				
				& \ebedwin   & \;\;0.1313 & $*$  &                & $*$ \\
				& \ptsadwin  & \;\;0.1462 & $*$  &                & $*$ \\
				& \spiritwin & \;\;0.7032 & $*$  &                & $*$ \\
				& \asedwin   & \;\;0.5470 & $*$  & $*$            & $*$ \\
				& \maedwin   & \;\;0.6670 &      &                & $*$ \\
				
				& \ebedwout   & \;\;0.2846 & $*$ &                &  \\
				& \ptsadwout  & \;\;0.2118 & $*$ &                & $*$ \\
				& \spiritwout & \;\;0.4563 &           &                & $*$ \\
				& \asedwout   & \;\;0.0580 & $*$ &                &  \\
				& \maedwout   & \;\;0.7328 &           & $*$      & $*$ \\
				
				& \ebeduin   & \;\;0.0892 & $*$  &                &  \\
				& \ptsaduin  & \;\;0.1607 & $*$  &                & $*$ \\
				& \spirituin & \;\;0.3996 & $*$  &                & $*$ \\
				& \aseduin   & \;\;0.1395 &            & $*$      & $*$ \\
				& \maeduin   & \;\;0.4439 &            & $*$      & $*$ \\
				
				& \ebeduout   & \;\;0.0225 & $*$ &           &  \\
				& \ptsaduout  & \;\;0.2546 &           &           & $*$ \\
				& \spirituout & \;\;0.1012 & $*$ &           & $*$ \\
				& \aseduout   & \;\;0.0870 & $*$ &           &  \\
				& \maeduout   & \;\;0.4181 &           &           & $*$ \\
				
				\hline \hline
				\parbox[t]{2mm}{\multirow{7}{*}{\rotatebox[origin=c]{90}{{\em Consensus}}}} 
				& \IR          & $*$\;{0.7121} & $*$\;{0.5660}  & \;\;\;0.6577         & $*$\;{0.7496} \\    
				& \KY	      & $*$\;{0.3033} & $*$\;{0.2495}  & \;\;\;0.5361         & $*$\;{0.5066} \\    
				& \rra	      & $*$\;{0.5948} & $*$\;{0.5348}  & \;\;\;0.4948    	  & $*$\;{0.5774} \\    
				& \uniavg   & $*$\;{0.4838} & $*$\;{0.4325}  & $*$\;{0.6047}        & $*$\;{0.5336} \\    
				& \unimax   & $*$\;{0.3020} & $*$\;{0.2242}  & \;\;\;0.6633    	  & $*$\;{0.4280} \\    
				& \mmavg    & $*$\;{0.5673} & $*$\;{0.4662}  & \;\;\;0.6761         & $*$\;{0.7217} \\    
				& \mmmax    & $*$\;{0.0216} & $*$\;{0.0216}  & $*$\;{0.5355}        & \;\;\;0.0222 \\
				\hline \hline    
				\multicolumn{2}{c} {Final Ensemble} \vline
				& \;\;\;0.5420 &	\;\;\;0.4697 &	\;\;\;0.7018 &	\;\;\;\textbf{0.7798} \\
				\hline    
			\end{tabular}
		}
		\end{center}
	}}
	\vspace{-0.25in}
	\label{tab:Enron_20}
\end{table}



\begin{table}[H]
	\caption{\small Accuracy of ensembles for RealityMining Voice Call (directed) (10 components) (features: weighted in-/out-degree). \;\;
	$*$: selected detector/consensus results.}
	\vspace{-0.25in}
	{\small {
			\begin{center}
				\resizebox{\columnwidth}{!}{%
				\begin{tabular}{c|l|c|c|c|c}
					\hline	
					&       & \fulle            & \dive            & \selectv        & \selecth        \\
					\hline\hline
					\parbox[t]{2mm}{\multirow{10}{*}{\rotatebox[origin=c]{90}{\textit{Base Algorithms}}}}
					
					& \ebedwin   & \;\;0.3508  & $*$       &           &  \\
					& \ptsadwin  & \;\;0.6284  &           &           & $*$ \\
					& \spiritwin & \;\;0.8309  & $*$       &           & $*$ \\
					& \asedwin   & \;\;0.9437  &           & $*$       & $*$ \\
					& \maedwin   & \;\;0.8809  & $*$       &           & $*$ \\	
					
					& \ebedwout   & \;\;0.4122  & $*$       &           &  \\
					& \ptsadwout  & \;\;0.6273  &           &           & $*$ \\
					& \spiritwout & \;\;0.7346  &           & $*$       & $*$ \\
					& \asedwout   & \;\;0.9500  &           &           & $*$ \\
					& \maedwout   & \;\;0.8758  &           &           & $*$ \\

					\hline
					\parbox[t]{2mm}{\multirow{7}{*}{\rotatebox[origin=c]{90}{{\em Consensus}}}} 
					& \IR          &$*$\;0.7544      & \;\;\;0.6169       & \;\;\;0.8880           & $*$\;0.8222     \\    
					& \KY	      &$*$\;0.8221      & $*$\;0.7708        & \;\;\;0.8619           & $*$\;0.9309     \\    
					& \rra	      &$*$\;0.8154      & \;\;\;0.5936       & \;\;\;0.8901           & $*$\;0.9416     \\    
					& \uniavg   &$*$\;0.7798      & $*$\;0.6413        & $*$\;0.8370            & $*$\;0.9098     \\    
					& \unimax   &$*$\;0.6704      & \;\;\;0.5757       & \;\;\;0.7786       	  & $*$\;0.7833     \\    
					& \mmavg    &$*$\;0.9190      & $*$\;0.9162        & \;\;\;0.8835           & $*$\;0.9183     \\    
					& \mmmax    &$*$\;0.4380      & $*$\;0.8934        & \;\;\;0.7569           & \;\;\;0.4380     \\
					\hline \hline  
				
					\multicolumn{2}{c} {\normalsize Final Ensemble} \vline
					& \;\;\;0.7302 &	\;\;\;0.8724 &	\;\;\;0.8370	& \;\;\;\textbf{0.9045} \\
					\hline    
				\end{tabular}
			}
			\end{center}
			}}  
		\vspace{-0.25in}
		\label{tab:Voice_Call_10}
	\end{table}


\begin{table}[H]
\caption{\small Accuracy of ensembles for RealityMining Voice Call (directed) (20 components)
(features: weighted in-/out-degree and unweighted in-/out-degree)
$*$: selected detector/consensus results.}
\vspace{-0.25in}
{\small {
		\begin{center}
			\resizebox{\columnwidth}{!}{%
			\begin{tabular}{c|l|c|c|c|c}
				\hline	
				&       & \fulle            & \dive            & \selectv        & \selecth        \\
				\hline\hline
				\parbox[t]{2mm}{\multirow{20}{*}{\rotatebox[origin=c]{90}{\textit{Base Algorithms}}}}
				
				& \ebedwin   & \;\;0.3508  & $*$       &           &  \\
				& \ptsadwin  & \;\;0.6284  &           &           & $*$ \\
				& \spiritwin & \;\;0.8309  &           & $*$       & $*$ \\
				& \asedwin   & \;\;0.9437  &           & $*$       & $*$ \\
				& \maedwin   & \;\;0.8809  & $*$       &           & $*$ \\	
				
				& \ebedwout   & \;\;0.4122  & $*$       &           &  \\
				& \ptsadwout  & \;\;0.6273  &           &           & $*$ \\
				& \spiritwout & \;\;0.7346  &           &           & $*$ \\
				& \asedwout   & \;\;0.9500  &           &           & $*$ \\
				& \maedwout   & \;\;0.8758  &           &           & $*$ \\
				
				& \ebeduin   & \;\;0.4173  &           &           &  \\
				& \ptsaduin  & \;\;0.8636  & $*$       &           & $*$ \\
				& \spirituin & \;\;0.8313  &           &           & $*$ \\
				& \aseduin   & \;\;0.9191  &           &           & $*$ \\
				& \maeduin   & \;\;0.8706  & $*$       &           & $*$ \\
				
				& \ebeduout   & \;\;0.4800  &           &           & $*$ \\
				& \ptsaduout  & \;\;0.8665  &           &           & $*$ \\
				& \spirituout & \;\;0.7480  &           &           & $*$ \\
				& \aseduout   & \;\;0.9229  & $*$       &           & $*$ \\
				& \maeduout   & \;\;0.9115  &           &           & $*$ \\
				
				\hline 
				\parbox[t]{2mm}{\multirow{7}{*}{\rotatebox[origin=c]{90}{{\em Consensus}}}} 
				& \IR          &$*$\;0.8035  & \;\;\;0.7952    & \;\;\;0.9240       & $*$\;0.8681 \\    
				& \KY	      &$*$\;0.9064  & \;\;\;0.9018    & \;\;\;0.9076       & $*$\;0.9158 \\    
				& \rra	      &$*$\;0.8866  & $*$\;0.7771     & \;\;\;0.9013       & $*$\;0.9311 \\    
				& \uniavg   &$*$\;0.8598  & \;\;\;0.9192    & $*$\;0.8448        & $*$\;0.9102 \\    
				& \unimax   &$*$\;0.6844  & $*$\;0.6863     & \;\;\;0.8517   	   & $*$\;0.7611 \\    
				& \mmavg    &$*$\;0.9321  & $*$\;0.9083     & $*$\;0.8312        & $*$\;0.9134 \\    
				& \mmmax    &$*$\;0.4380  & $*$\;0.8858     & \;\;\;0.8015       & \;\;\;0.4380     \\
				\hline \hline    
				\multicolumn{2}{c} {Final Ensemble} \vline
				& \;\;\;0.8011 &	\;\;\;0.8335 &	\;\;\;0.8847 &	\;\;\;\textbf{0.8949} \\
				\hline    
			\end{tabular}
		}
		\end{center}
	}}
	\vspace{-0.25in}
	\label{tab:Voice_Call_20}
\end{table}

\begin{table}[H]
\caption{\small Accuracy of ensembles for RealityMining Bluetooth (undirected) (10 components)
(feature: weighted and unweighted degree).
$*$: selected detector/consensus results.}
\vspace{-0.25in}
{\small {
		\begin{center}
			\resizebox{\columnwidth}{!}{%
			\begin{tabular}{c|l|c|c|c|c}
				\hline	
				&       & \fulle            & \dive            & \selectv        &  \selecth  \\
				\hline\hline
				\parbox[t]{2mm}{\multirow{10}{*}{\rotatebox[origin=c]{90}{\textit{Base Algorithms}}}}
				& \ebedwdeg   & \;\;0.4363 & $*$ &           &   \\
				& \ptsadwdeg  & \;\;0.5820 & $*$ &           & $*$ \\
				& \spiritwdeg & \;\;0.9499 & $*$ &           & $*$ \\
				& \asedwdeg   & \;\;0.8601 &           & $*$ & $*$ \\
				& \maedwdeg   & \;\;0.8359 & $*$ &           & $*$ \\
				
				& \ebedudeg   & \;\;0.4966 & $*$ &           &   \\
				& \ptsadudeg  & \;\;0.8694 &           & $*$ & $*$\\
				& \spiritudeg & \;\;0.9162 &           & $*$ & $*$ \\
				& \asedudeg   & \;\;0.7662 &           & $*$ & $*$ \\
				& \maedudeg   & \;\;0.8788 & $*$ &           & $*$ \\
				\hline \hline
				
				\parbox[t]{2mm}{\multirow{7}{*}{\rotatebox[origin=c]{90}{{\em Consensus}}}} 
				& \IR            & $*$\;0.8646 & $*$\;0.8255    & \;\;\;0.8790          & $*$\;{0.8538} \\    
				& \KY	        & $*$\;0.9534 & \;\;\;0.9169   & \;\;\;0.9698          & $*$\;{0.9361} \\    
				& \rra	        & $*$\;0.9413 & \;\;\;0.8318   & \;\;\;0.9693    	   & $*$\;{0.9684} \\    
				& \uniavg     & $*$\;0.9071 & \;\;\;0.8654   & $*$\;0.9193           & $*$\;{0.9225} \\    
				& \unimax     & $*$\;0.6973 & $*$\;0.6122    & \;\;\;0.8270    	   & $*$\;{0.7126} \\    
				& \mmavg      & $*$\;0.9407 & $*$\;0.9340    & \;\;\;0.8596          & $*$\;{0.8892} \\    
				& \mmmax      & $*$\;0.6461 & $*$\;0.6374    & \;\;\;0.8830          & \;\;\;0.6461 \\
				\hline \hline    
				\multicolumn{2}{c} {Final Ensemble} \vline
				& \;\;\;0.8398 &	\;\;\;0.7735 &	\;\;\;\textbf{0.9193} &   \;\;\;0.8886 \\
				\hline    
			\end{tabular}
		}
		\end{center}
	}}
	\vspace{-0.25in}
	\label{tab:Bluetooth_10}
\end{table}

\begin{table}[h]
\caption{\small Accuracy of ensembles for RealityMining SMS (directed) (10 components)
(features: weighted in-/out-degree).
$*$: selected detector/consensus results.}
\vspace{-0.25in}
{\small {
		\begin{center}
			\resizebox{\columnwidth}{!}{%
			\begin{tabular}{c|l|c|c|c|c}
				\hline	
				&       & \fulle            & \dive            & \selectv        & \selecth        \\
				\hline\hline
				\parbox[t]{2mm}{\multirow{10}{*}{\rotatebox[origin=c]{90}{\textit{Base Algorithms}}}}
				
				& \ebedwin   & \;\;0.6117 & $*$       &           &  \\
				& \ptsadwin  & \;\;0.7003 &           &           & $*$\\
				& \spiritwin & \;\;0.9256 &           & $*$       & $*$ \\
				& \asedwin   & \;\;0.6338 & $*$       &           & $*$ \\
				& \maedwin   & \;\;0.9002 & $*$       &           & $*$ \\
				
				& \ebedwout   & \;\;0.5595 & $*$       &           &  \\
				& \ptsadwout  & \;\;0.7023 &           & $*$       & $*$   \\
				& \spiritwout & \;\;0.8656 & $*$       &           & $*$ \\
				& \asedwout   & \;\;0.9102 &           & $*$       & $*$ \\
				& \maedwout   & \;\;0.9259 &           & $*$       & $*$ \\

				\hline \hline
				\parbox[t]{2mm}{\multirow{7}{*}{\rotatebox[origin=c]{90}{{\em Consensus}}}} 
				& \IR          & $*$\;0.8309     & \;\;\;0.8174     & \;\;\;0.8933         & $*$\;0.8044  \\    
				& \KY	      & $*$\;0.9491     & $*$\;0.8779      & \;\;\;0.9511         & $*$\;0.9386  \\    
				& \rra	      & $*$\;0.8761     & $*$\;0.8424      & \;\;\;0.9578         & $*$\;0.9516  \\    
				& \uniavg   & $*$\;0.8531     & \;\;\;0.8247     & $*$\;0.9283          & $*$\;0.8684  \\    
				& \unimax   & $*$\;0.8205     & $*$\;0.7632      & \;\;\;0.8829         & $*$\;0.8678  \\    
				& \mmavg    & $*$\;0.9276     & $*$\;0.9487      & \;\;\;0.9492         & $*$\;0.9084  \\    
				& \mmmax    & $*$\;0.8907     & $*$\;0.8577      & \;\;\;0.9410         & \;\;\;0.9011     \\
				\hline \hline    
				\multicolumn{2}{c} {Final Ensemble} \vline
				& \;\;\;0.9092 &	\;\;\;0.8598 &	\;\;\;\textbf{0.9283}	 & \;\;\;0.9217 \\
				\hline    
			\end{tabular}
		}
		\end{center}
	}}
	\vspace{-0.25in}
	\label{tab:SMS_10}
\end{table}

\begin{table}[H]
\caption{\small Accuracy of ensembles for RealityMining SMS (directed) (20 components)
(features: weighted in-/out-degree and unweighted in-/out-degree).
$*$: selected detector/consensus results.}
\vspace{-0.1in}
{\small {
		\begin{center}
			\resizebox{\columnwidth}{!}{%
			\begin{tabular}{c|l|c|c|c|c}
				\hline	
				&       & \fulle            & \dive            & \selectv        & \selecth        \\
				\hline\hline
				\parbox[t]{2mm}{\multirow{20}{*}{\rotatebox[origin=c]{90}{\textit{Base Algorithms}}}}
				
				& \ebedwin   & \;\;0.6117  & $*$       &           & $*$ \\
				& \ptsadwin  & \;\;0.7003  &           &           & $*$\\
				& \spiritwin & \;\;0.9256 &           &           & $*$ \\
				& \asedwin   & \;\;0.6338 & $*$       &           & $*$ \\
				& \maedwin   & \;\;0.9002 &           &           & $*$ \\
				
				& \ebedwout   & \;\;0.5595  &           &           & $*$ \\
				& \ptsadwout  & \;\;0.7023 &           &           &     \\
				& \spiritwout & \;\;0.8656 &           &           & $*$ \\
				& \asedwout   & \;\;0.9102 &           &           & $*$ \\
				& \maedwout   & \;\;0.9259 &           & $*$       & $*$ \\
				
				& \ebeduin   & \;\;0.4407 & $*$       &           &  \\
				& \ptsaduin  & \;\;0.7809 & $*$       &           & $*$ \\
				& \spirituin & \;\;0.7841 &           &           & $*$ \\
				& \aseduin   & \;\;0.6248 & $*$       &           & $*$ \\
				& \maeduin   & \;\;0.8297 & $*$       &           & $*$ \\
				
				& \ebeduout   & \;\;0.3246 &           &           &          \\
				& \ptsaduout  & \;\;0.9157 &           & $*$       & $*$ \\
				& \spirituout & \;\;0.8744 &           &           & $*$ \\
				& \aseduout   & \;\;0.9150 &           &           & $*$ \\
				& \maeduout   & \;\;0.8005 &           &           & $*$ \\
				
				\hline \hline
				\parbox[t]{2mm}{\multirow{7}{*}{\rotatebox[origin=c]{90}{{\em Consensus}}}} 
				& \IR       & $*$\;0.9135     & \;\;\;0.6751     & \;\;\;0.9634         & $*$\;0.9230     \\    
				& \KY	    & $*$\;0.9286     & $*$\;0.7567      & \;\;\;0.9094         & $*$\;0.9325     \\    
				& \rra	    & $*$\;0.9568     & \;\;\;0.6465     & \;\;\;0.9418         & $*$\;0.9583     \\    
				& \uniavg   & $*$\;0.8791     & \;\;\;0.6499     & $*$\;0.9294          & $*$\;0.9156     \\    
				& \unimax   & $*$\;0.7173     & $*$\;0.6696      & \;\;\;0.9342         & \;\;\;0.8650              \\    
				& \mmavg    & $*$\;0.9107     & $*$\;0.8942      & \;\;\;0.8519         & $*$\;0.9138     \\    
				& \mmmax    & $*$\;0.8895     & $*$\;0.8480      & \;\;\;0.9307         & \;\;\;0.8895          \\
				\hline \hline    
				\multicolumn{2}{c} {Final Ensemble} \vline
				& \;\;\;0.9542 &	\;\;\;0.8749 &	\;\;\;0.9294 & \;\;\;\textbf{0.9621} \\
				\hline    
			\end{tabular}
		}
		\end{center}
	}}
	\vspace{-0.25in}
	\label{tab:SMS_20}
\end{table}

\begin{table}[H]
	\caption{\small Accuracy of ensembles for TwitterSecurity (undirected) (10 components)
	(features: weighted and unweighted degree).
	$*$: selected detector/consensus results.}
	\vspace{-0.25in}
	{\small {
			\begin{center}
				\resizebox{\columnwidth}{!}{%
				\begin{tabular}{c|l|c|c|c|c}
					\hline	
					&       & \fulle            & \dive            & \selectv        & \selecth        \\
					\hline\hline
					\parbox[t]{2mm}{\multirow{10}{*}{\rotatebox[origin=c]{90}{\textit{Base Algorithms}}}}
					
					& \ebedwdeg   & \;\;0.4000     & $*$       &           & $*$ \\
					& \ptsadwdeg  & \;\;0.5400     & $*$       &           & $*$ \\
					& \spiritwdeg & \;\;0.4467     & $*$       & $*$ & $*$ \\
					& \asedwdeg   & \;\;0.6200     & $*$       & $*$ & $*$ \\
					& \maedwdeg   & \;\;0.4933     &                 &           & $*$ \\
					
					& \ebedudeg   & \;\;0.4133     & $*$       & $*$ & $*$ \\
					& \ptsadudeg  & \;\;0.5467     &                 & $*$ & $*$ \\
					& \spiritudeg & \;\;0.3867     & $*$       &           & $*$ \\
					& \asedudeg   & \;\;0.5400     & $*$       &           & $*$ \\
					& \maedudeg   & \;\;0.4533     & $*$       &           &     \\

					\hline \hline
					\parbox[t]{2mm}{\multirow{7}{*}{\rotatebox[origin=c]{90}{{\em Consensus}}}} 
					& \IR       & $*$\;0.4467     & \;\;\;0.4267        & \;\;\;0.5133             & $*$\;0.4667   \\    
					& \KY	    & $*$\;0.5667     & \;\;\;0.5333        & \;\;\;0.5333             & \;\;\;0.5800      \\    
					& \rra	    & $*$\;0.5867     & $*$\;0.5333         & \;\;\;0.5467               & $*$\;0.5933  \\    
					& \uniavg   & $*$\;0.5600     & \;\;\;0.5000        & $*$\;0.5467              & $*$\;0.6000   \\    
					& \unimax   & $*$\;0.4533     & $*$\;0.4400         & \;\;\;0.5800               & \;\;\;0.4533      \\    
					& \mmavg    & $*$\;0.5333     & $*$\;0.5667         & \;\;\;0.5267               & \;\;\;0.5600      \\    
					& \mmmax    & $*$\;0.3667     & $*$\;0.3667         & \;\;\;0.5533               & \;\;\;0.5733  \\
					\hline \hline    
					\multicolumn{2}{c} {Final Ensemble} \vline
					& \;\;\;0.5200 & \;\;\;0.4800 &	\;\;\;0.5467 & \;\;\;\textbf{0.5867} \\
					\hline    
				\end{tabular}
			}
			\end{center}
		}}
		\vspace{-0.25in}
		\label{tab:twitter_2014_DSandT_10}
	\end{table}



\begin{table}[H]
	\caption{\small Challenge Network: (Feature: unweighted in \& outdegree)}
	\vspace{-0.1in}
	{\small {
			\begin{center}
			    \resizebox{\columnwidth}{!}{%
				\begin{tabular}{c|l|c|c|c|c}
					\hline	
					& \algorithms      & \fulle            & \dive            & \selectv        & \selecth        \\
					\hline\hline
					\parbox[t]{2mm}{\multirow{10}{*}{\rotatebox[origin=c]{90}{\textit{Base Algorithms}}}}
					
					& \ebeduin   & \;\;0.1587     & $*$       &                & $*$ \\
					& \ptsaduin  & \;\;0.5208     & $*$       &                & $*$ \\
					& \spirituin & \;\;0.7556     & $*$       & $*$   & $*$ \\
					& \aseduin   & \;\;1.0000     & $*$       & $*$   & $*$ \\
					& \maeduin   & \;\;1.0000     &                    & $*$   & $*$ \\
					
					& \ebeduout   & \;\;0.7333     & $*$       &                & $*$ \\
					& \ptsaduout  & \;\;0.0994     & $*$       &                & $*$ \\
					& \spirituout & \;\;0.4021     & $*$       & $*$   & $*$ \\
					& \aseduout   & \;\;1.0000     & $*$       & $*$   & $*$ \\
					& \maeduout   & \;\;1.0000     & $*$       & $*$   & $*$  \\

					\hline \hline
					\parbox[t]{2mm}{\multirow{7}{*}{\rotatebox[origin=c]{90}{{\em Consensus}}}} 
					& \IR          & $*$\;1.0000     & $*$\;1.0000   & $*$\;1.0000      & $*$\;1.0000    \\    
					& \KY	      & $*$\;0.9167     & \;\;\;0.9167  & $*$\;1.0000      & $*$\;0.9167      \\    
					& \rra	      & $*$\;0.9167     & $*$\;0.9167   & \;\;\;1.0000     & $*$\;0.9167  \\    
					& \uniavg   & $*$\;1.0000     & $*$\;1.0000   & $*$\;1.0000      & $*$\;1.0000   \\    
					& \unimax   & $*$\;0.2778     & $*$\;0.2778   & \;\;\;1.0000     & \;\;\;0.2778                \\    
					& \mmavg    & $*$\;1.0000     & $*$\;1.0000   & $*$\;1.0000      & $*$\;1.0000      \\    
					& \mmmax    & $*$\;0.7000     & \;\;\;0.7000  & \;\;\;1.0000     & $*$\;0.7000  \\
					\hline \hline    
					\multicolumn{2}{c} {Final Ensemble} \vline
					& \;\;\;1.0000 &   \;\;\;1.0000 & \;\;\;1.0000 & \;\;\;1.0000 \\
					\hline    
				\end{tabular}
			}
			\end{center}
		}}
		\vspace{-0.25in}
		\label{tab:challenge_network_10}
	\end{table}
	
	
\begin{table}[H]
\caption{\small Ensemble results for Challenge Network}
\vspace{-0.1in}
{\small {
		\begin{center} 
			\begin{tabular}{l|l|r} \hline
				Method  & AP & significance   \\ \hline \hline
				\fulle  & 1.0000     & n/a   \\ \hline
				\dive  & 1.0000      & n/a  \\ \hline
				(i) Random Ensemble  & 0.8831  &$\pm$0.1380  \\ 
				(6/10 + 4/7) & & \\ 
				\selectv & 1.0000       &$=\mu + 0.8471 \sigma$ \\ \hline
				(ii) Random Ensemble  & 0.9405 &$\pm$0.0407  \\ 
				(10/10 + 6/7) & & \\ 
				\selecth & 1.0000      &$=\mu + 1.4619 \sigma$   \\  
				\hline
			\end{tabular}
		\end{center}
	}}
	\vspace{-0.25in}
	\label{big_table_challenge_network_10}
\end{table}


\begin{table}[H]
	\caption{\small Challenge Network: Characterization of time tick 377 (Feature: unweighted in \& outdegree)}
	\vspace{-0.1in}
	{\small {
			\begin{center}
				\resizebox{\columnwidth}{!}{%
					\begin{tabular}{c|l|c|c|c|c}
						\hline	
						& \algorithms      & \fulle            & \dive            & \selectv        & \selecth        \\
						\hline\hline
						\parbox[t]{2mm}{\multirow{10}{*}{\rotatebox[origin=c]{90}{\textit{Base Algorithms}}}}
						
						& \ebeduin   & \;\;1.0000         & $*$       &           & $*$ \\
						& \ptsaduin  & \;\;0.2000         & $*$       &           & $*$ \\
						& \spirituin & \;\;0.5000         & $*$       &           & $*$ \\
						& \aseduin   & \;\;1.0000         &           & $*$       & $*$ \\
						& \maeduin   & \;\;1.0000         &           & $*$       & $*$ \\
						
						& \ebeduout   & \;\;0.2500        &           &           & $*$ \\
						& \ptsaduout  & \;\;0.2000        &           &           & $*$ \\
						& \spirituout & \;\;0.0270        & $*$       &           &     \\
						& \aseduout   & \;\;1.0000        & $*$       &           & $*$ \\
						& \maeduout   & \;\;1.0000        & $*$       & $*$       & $*$  \\
						
						\hline \hline
						\parbox[t]{2mm}{\multirow{7}{*}{\rotatebox[origin=c]{90}{{\em Consensus}}}} 
						& \IR       & $*$\;1.0000     & $*$\;1.0000   & $*$\;1.0000      & $*$\;1.0000     \\    
						& \KY	    & $*$\;1.0000     & $*$\;1.0000   & \;\;\;1.0000     & $*$\;1.0000     \\    
						& \rra	    & $*$\;1.0000     & $*$\;1.0000   & $*$\;1.0000      & $*$\;1.0000      \\    
						& \uniavg   & $*$\;1.0000     & $*$\;1.0000   & $*$\;1.0000      & $*$\;1.0000   \\    
						& \unimax   & $*$\;0.5000     & \;\;\;0.5000  & \;\;\;1.0000     & $*$\;0.5000    \\    
						& \mmavg    & $*$\;1.0000     & $*$\;1.0000   & \;\;\;0.2500     & $*$\;1.0000      \\    
						& \mmmax    & $*$\;0.0147     & $*$\;0.0200   & \;\;\;0.2500     & \;\;\;0.0147      \\
						\hline \hline    
						\multicolumn{2}{c} {Final Ensemble} \vline
						& \;\;\;1.0000 &   \;\;\;1.0000 & \;\;\;1.0000 & \;\;\;1.0000 \\
						\hline    
					\end{tabular}
				}
			\end{center}
		}}
		\vspace{-0.25in}
		\label{tab:challenge_network_characterization_10}
	\end{table}
		


\begin{table}[H]
\caption{\small Challenge Network: (Feature: weighted in \& outdegree and unweighted in \& outdegree)}
\vspace{-0.1in}
{\small {
		\begin{center}
			\resizebox{\columnwidth}{!}{%
				\begin{tabular}{c|l|c|c|c|c}
					\hline	
					& \algorithms      & \fulle            & \dive            & \selectv        & \selecth        \\
					\hline\hline
					\parbox[t]{2mm}{\multirow{20}{*}{\rotatebox[origin=c]{90}{\textit{Base Algorithms}}}}
					
					& \ebedwin   & \;\;0.0054     & $*$       &           & $*$ \\
					& \ptsadwin  & \;\;0.1183     & $*$       & $*$       & $*$\\
					& \spiritwin & \;\;0.0039     &           & $*$       & $*$ \\
					& \asedwin   & \;\;0.0178     & $*$       &           & $*$ \\
					& \maedwin   & \;\;0.0568     &           &           & $*$ \\
					
					& \ebedwout   & \;\;0.0064    & $*$       &           & $*$ \\
					& \ptsadwout  & \;\;0.6717    & $*$       & $*$       & $*$    \\
					& \spiritwout & \;\;0.0036    &           &           & $*$ \\
					& \asedwout   & \;\;0.0104    & $*$       & $*$       & $*$ \\
					& \maedwout   & \;\;0.0494    & $*$       &           & $*$ \\
					
					& \ebeduin   & \;\;0.1587     & $*$       &           & $*$ \\
					& \ptsaduin  & \;\;0.5208     & $*$       & $*$       & $*$ \\
					& \spirituin & \;\;0.7556     & $*$       &           & $*$ \\
					& \aseduin   & \;\;1.0000     & $*$       &           & $*$ \\
					& \maeduin   & \;\;1.0000     &           &           & $*$ \\
					
					& \ebeduout   & \;\;0.7333    & $*$       &           & $*$ \\
					& \ptsaduout  & \;\;0.0994    & $*$       &           & $*$ \\
					& \spirituout & \;\;0.4021    & $*$       &           & $*$ \\
					& \aseduout   & \;\;1.0000    & $*$       &           & $*$ \\
					& \maeduout   & \;\;1.0000    & $*$       &           & $*$  \\

					\hline \hline
					\parbox[t]{2mm}{\multirow{7}{*}{\rotatebox[origin=c]{90}{{\em Consensus}}}} 
					& \IR       & $*$\;0.8333     & $*$\;0.8333   & $*$\;0.4712          & $*$\;0.8333        \\    
					& \KY	    & $*$\;0.4720     & $*$\;0.5167   & $*$\;0.1730          & $*$\;0.4720          \\    
					& \rra	    & $*$\;0.8095     & $*$\;0.7778   & $*$\;0.3827          & $*$\;0.8095      \\    
					& \uniavg   & $*$\;0.5000     & $*$\;0.6250   & $*$\;0.0956          & $*$\;0.0339   \\    
					& \unimax   & $*$\;0.2778     & $*$\;0.2778   & $*$\;0.3503          & $*$\;0.0219                    \\    
					& \mmavg    & $*$\;0.0244     & $*$\;0.3432   & $*$\;0.0093          & $*$\;0.0244          \\    
					& \mmmax    & $*$\;0.0151     & $*$\;0.0151   & $*$\;0.0499          & \;\;\;0.0151      \\
					\hline \hline    
					\multicolumn{2}{c} {Final Ensemble} \vline
					& \;\;\;0.6325 &   \;\;\;0.6667 & \;\;\;0.3575 & \;\;\;0.7500 \\
					\hline    
				\end{tabular}
			}
		\end{center}
	}}
	\vspace{-0.25in}
	\label{tab:challenge_network_20}
\end{table}


\begin{table}[H]
\caption{\small Ensemble results for Challenge Network}
\vspace{-0.1in}
{\small {
		\begin{center} 
			\begin{tabular}{l|l|r} \hline
				Method  & AP & significance   \\ \hline \hline
				\fulle & 0.6325     & n/a   \\ \hline
				\dive  & 0.6667      & n/a  \\ \hline
				(i) Random Ensemble  & 0.5975   &$\pm$0.2599  \\ 
				(5/20 + 7/7) & & \\ 
				\selectv & 0.3575       &$=\mu - 0.9234 \sigma$ \\ \hline
				(ii) Random Ensemble  & 0.6128  &$\pm$0.0881  \\ 
				(20/20 + 6/7) & & \\ 
				\selecth & 0.7500      &$=\mu + 1.5573 \sigma$   \\  
				\hline
			\end{tabular}
		\end{center}
	}}
	\vspace{-0.25in}
	\label{big_table_challenge_network_20}
\end{table}


\begin{table}[H]
\caption{\small Challenge Network: Characterization of time tick 377 (Feature: weighted in \& outdegree and unweighted in \& outdegree)}
\vspace{-0.1in}
{\small {
		\begin{center}
			\resizebox{\columnwidth}{!}{%
				\begin{tabular}{c|l|c|c|c|c}
					\hline	
					& \algorithms      & \fulle            & \dive            & \selectv        & \selecth        \\
					\hline\hline
					\parbox[t]{2mm}{\multirow{20}{*}{\rotatebox[origin=c]{90}{\textit{Base Algorithms}}}}
					
					& \ebedwin   & \;\;0.0909         & $*$       & $*$       & $*$ \\
					& \ptsadwin  & \;\;1.0000         & $*$       &           & $*$\\
					& \spiritwin & \;\;0.0238         &           &           &  \\
					& \asedwin   & \;\;0.3333         &           & $*$       & $*$ \\
					& \maedwin   & \;\;1.0000         & $*$       &           & $*$ \\
					
					& \ebedwout   & \;\;0.0385        &           &           &     \\
					& \ptsadwout  & \;\;0.1429        & $*$       &           & $*$    \\
					& \spiritwout & \;\;0.0133        & $*$       &           &  \\
					& \asedwout   & \;\;0.2500        & $*$       & $*$       & $*$ \\
					& \maedwout   & \;\;1.0000        & $*$       &           & $*$ \\
					
					& \ebeduin   & \;\;1.0000         & $*$       &           & $*$ \\
					& \ptsaduin  & \;\;0.2000         & $*$       & $*$       & $*$ \\
					& \spirituin & \;\;0.5000         & $*$       &           &     \\
					& \aseduin   & \;\;1.0000         &           &           & $*$ \\
					& \maeduin   & \;\;1.0000         &           & $*$       & $*$ \\
					
					& \ebeduout   & \;\;0.2500        &           & $*$       & $*$ \\
					& \ptsaduout  & \;\;0.2000        &           &           & $*$ \\
					& \spirituout & \;\;0.0270        & $*$       &           &     \\
					& \aseduout   & \;\;1.0000        & $*$       &           & $*$ \\
					& \maeduout   & \;\;1.0000        &           &           & $*$  \\

					\hline \hline
					\parbox[t]{2mm}{\multirow{7}{*}{\rotatebox[origin=c]{90}{{\em Consensus}}}} 
					& \IR       & $*$\;1.0000     & $*$\;1.0000   & \;\;\;1.0000     & $*$\;1.0000     \\    
					& \KY	    & $*$\;1.0000     & $*$\;1.0000   & $*$\;1.0000      & $*$\;1.0000     \\    
					& \rra	    & $*$\;1.0000     & $*$\;1.0000   & \;\;\;1.0000     & $*$\;1.0000      \\    
					& \uniavg   & $*$\;1.0000     & $*$\;1.0000   & $*$\;1.0000      & $*$\;1.0000   \\    
					& \unimax   & $*$\;0.1667     & $*$\;0.2000   & \;\;\;0.5000     & \;\;\;0.3333    \\    
					& \mmavg    & $*$\;1.0000     & \;\;\;1.0000  & \;\;\;1.0000     & $*$\;1.0000      \\    
					& \mmmax    & $*$\;0.0128     & $*$\;0.0128   & \;\;\;0.0189     & \;\;\;0.0135      \\
					\hline \hline    
					\multicolumn{2}{c} {Final Ensemble} \vline
					& \;\;\;1.0000 &   \;\;\;1.0000 & \;\;\;1.0000 & \;\;\;1.0000 \\
					\hline    
				\end{tabular}
			}
		\end{center}
	}}
	\vspace{-0.25in}
	\label{tab:challenge_network_characterization_20}
\end{table}
%
%
%

\begin{figure}[H]
\centering
\begin{tabular}{cc}
\hspace{-0.1in}\includegraphics[width=1.65in,height=1.35in]{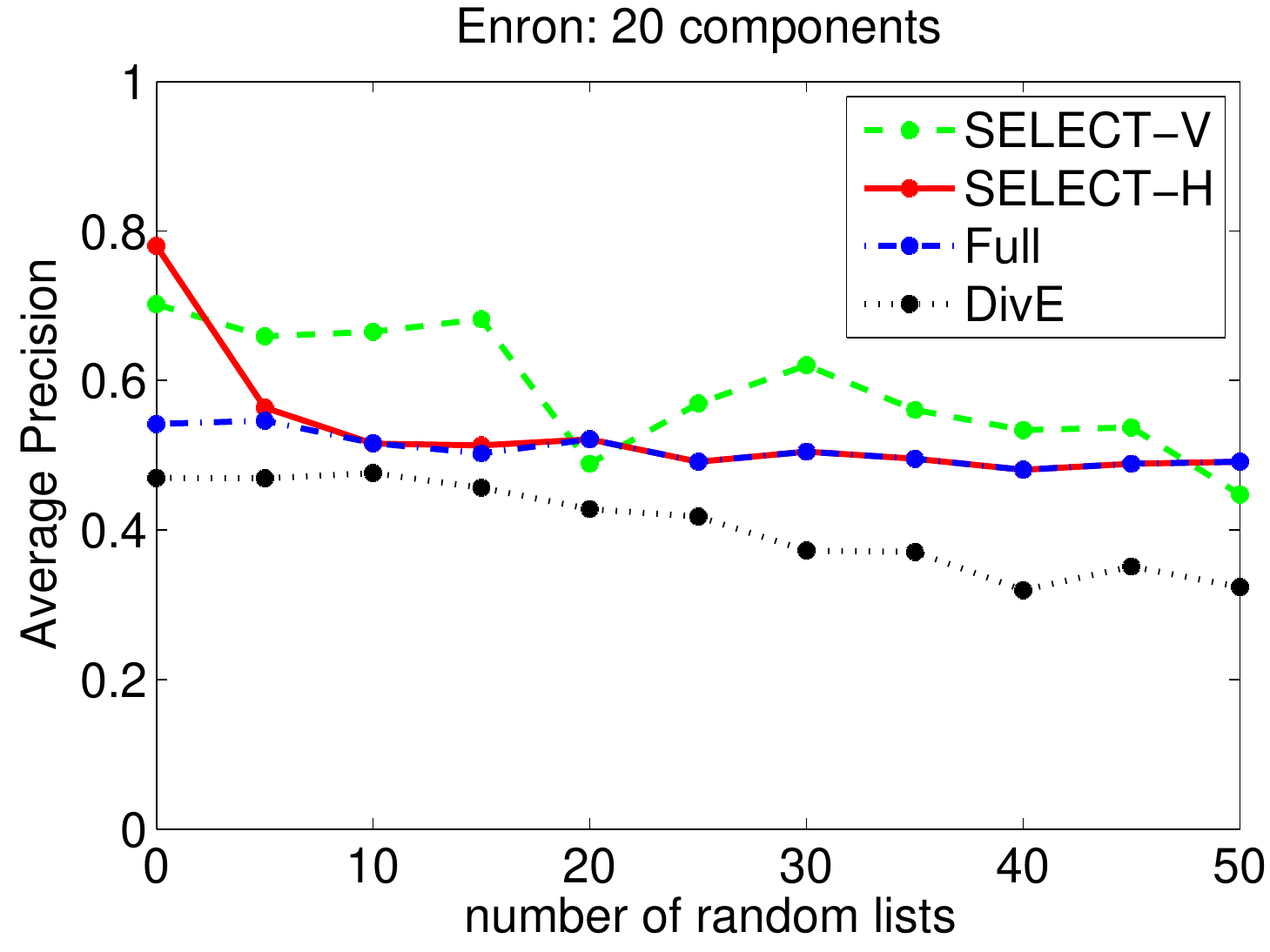} &
\hspace{-0.1in}\includegraphics[width=1.65in,height=1.35in]{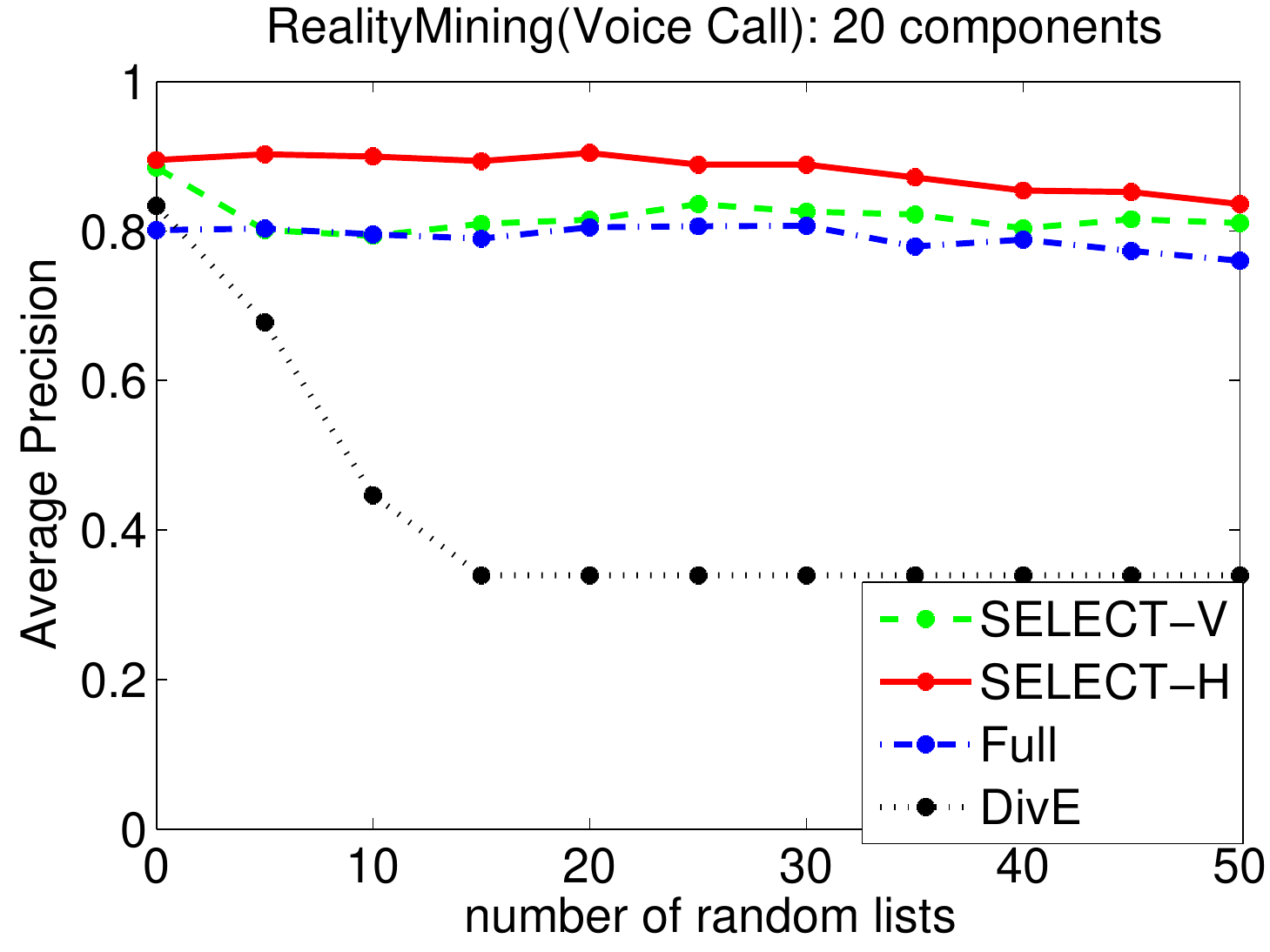} \\
  (a) EnronInc. & (b) RealityMining VC  \\\\
\multicolumn{2}{c}{\hspace{-0.1in}\includegraphics[width=1.65in,height=1.35in]{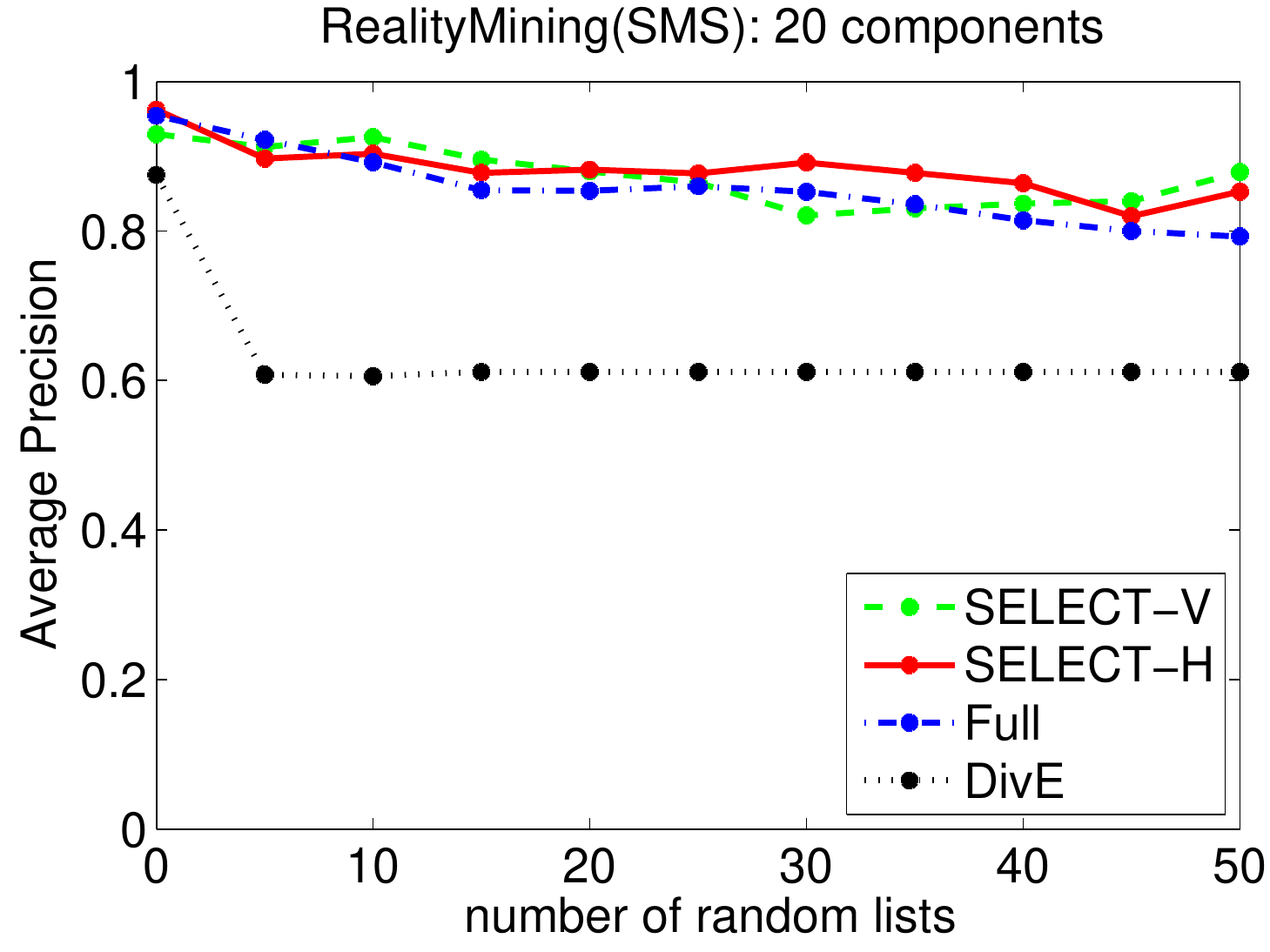} }
  \\
\multicolumn{2}{c}{(c) RealityMining SMS} \\
\end{tabular}
\vspace{-0.05in}
\caption{\small Analysis of accuracy when increasing number of random base results are introduced for ensembles with 20 components. Decline in accuracy under noise is most prominent for \dive.} 
\label{fig:noise2}
\vspace{-0.2in}
\end{figure}

%
%


\bibliographystylelatex{abbrv}
\bibliographylatex{BIB/refsmeas}


\end{document}